\documentclass[1p,number,sort&compress]{elsarticle}

\usepackage{amsfonts} 
\usepackage{amssymb} 
\usepackage{amsmath} 
\usepackage{subfigure} 
\usepackage{array} 
\usepackage{dcolumn} 
\usepackage{bm} 
\usepackage{latexsym} 
\usepackage{longtable} 
\graphicspath{{./figs/}}

\journal{Nuclear Physics B}

\begin{document}
\begin{frontmatter}


\title{ Covariance fitting of highly correlated data in lattice QCD }
%
\author[a]{Yong-Chull Jang}
\author[b]{Chulwoo Jung}
\ead{chulwoo@bnl.gov}
\author[a,c]{Weonjong Lee\corref{cor1}}
\ead{wlee@snu.ac.kr}
\author[a]{Boram Yoon\corref{cor2}}
%
%
\address[a]{
  Lattice Gauge Theory Research Center, FPRD, and CTP,\\
  Department of Physics and Astronomy, \\
  Seoul National University, Seoul, 151-747, South Korea
}
\address[b]{
  Physics Department, Brookhaven National Laboratory,
  Upton, NY11973, USA
}
\address[c]{
  Physics Department,
  University of Washington,
  Seattle, WA 98195-1560, USA
}
\cortext[cor1]{Corresponding author}
\cortext[cor2]{Principal corresponding author}
\date{\today}
\begin{abstract}
  We address a frequently asked question on the covariance fitting of
  the highly correlated data such as our $B_K$ data based on the SU(2)
  staggered chiral perturbation theory.
  Basically, the essence of the problem is that we do not have an
  accurate fitting function enough to fit extremely precise data.
  When eigenvalues of the covariance matrix are small, even a tiny
  error of fitting function yields large chi-square and spoils the
  fitting procedure.
  We have applied a number of prescriptions available in the market 
  such as the cut-off method, modified covariance matrix method, and
  Bayesian method. 
  We also propose a brand new method, the eigenmode shift method which
  allows a full covariance fitting without modifying the covariance
  matrix at all.
  In our case, the eigenmode shift (ES) method and Bayesian method turn
  out to be the best prescription to the problem.
  We also provide a pedagogical example of data analysis in which the
  diagonal approximation and the cut-off method fail in fitting
  manifestly, but the ES method and the Bayesian approach work well.
\end{abstract}

\end{frontmatter}

\section{Introduction \label{sec:intr}}
We have reported results of $B_K$ calculated using improved staggered
fermions with $N_f=2+1$ flavors in Ref.~\cite{wlee-10-3,Bae:2011ff}.
We refer to Ref.~\cite{wlee-10-3} and \cite{Bae:2011ff} as SW-1 and
SW-2 afterwards.
In SW-1, we use three different lattice spacings to control the
discretization errors.
The dominant error in this result comes from uncertainty in
the matching factors.
One hidden uncertainty is that we use the diagonal approximation
(uncorrelated fitting) instead of the full covariance fitting in
SW-1.
In fact, one of the most frequently asked questions on SW-1 is
why we do the uncorrelated fitting ({i.e.} using the diagonal
approximation) instead of the full covariance fitting.
Here, we would like to address this issue on the covariance fitting
and the diagonal approximation.

A significant difficulty in fitting the highly correlated data has been
pointed out in the literature such as
Refs.~\cite{Thacker:1990bm,Drummond:1992pg,kilcup-1994-1,michael-1994-1,
Michael:1994sz}.
In the literature, they have proposed a number of prescriptions such
as diagonal approximation \cite{michael-1994-1}, modifying the
covariance matrix \cite{Michael:1994sz}, and cut-off method in the
popular name of singular value decomposition (SVD)
\cite{Thacker:1990bm,Drummond:1992pg,kilcup-1994-1}.
The weakness of these approaches is that all of these methods try to
modify the covariance matrix one way or another.
Hence, we lose the true meaning of $\chi^2$ and we do not know the
quality of fitting in this case.

Therefore, we propose a new method, the eigenmode shift (ES) method,
which does not modify the covariance matrix but only use our freedom
to modify the fitting functional form based on the Bayesian method.
It turns out that the ES method allows for a probability interpretation
of quality of fitting based on the Bayesian $\chi^2$
distribution.\footnote{This is different from the normal $\chi^2$
  distribution, which assumes the uniform prior information. We will
  address this issue when we discuss on the Bayesian method.}
An alternative approach is the orthodox Bayesian method.
In this approach, we add higher order terms to the fitting function
with proper constraints until it fits the data. This also
turns out to be another good solution to the problem.

The paper is organized as follows.
In Sec.~\ref{sec:cov-fit}, we review the covariance fitting process and
give a physical meaning of covariance matrix.
In Sec.~\ref{sec:bk-cov-mat}, we address the problem with small eigenvalues
of covariance matrix.
In Sec.~\ref{sec:pres}, we list the possible solutions to the problem and
discuss about the pros and cons. Here, we explain the ES method.
In Sec.~\ref{sec:err-cov-mat}, we give a theoretical background for
the pdf (probability distribution function) of the eigenvalues of the
covariance matrix.
In Sec.~\ref{sec:conclude}, we close with some concluding remarks.

\section{Review of covariance fitting}
\label{sec:cov-fit}
Let us consider $N$ samples of unbiased estimates of
quantity $y_i$ with $i=1,2,3,\ldots,D$.
Here, the data set is $\{y_i(n) | n=1,2,3,\ldots,N \}$.
Let us assume that the samples $y_i(n)$ are statistically independent
in $n$ for fixed $i$ but are substantially correlated in $i$.
For example, a similar situation occurs in lattice gauge theory
calculations where there are $N$ independent gauge configurations and
$B_K$ values with multiple choices of different quark mass pairs of
$m_x$ (valence down quark mass) and $m_y$ (valence strange quark
mass), which corresponds to $D$ Green functions measured over the
gauge configurations.
An introduction to this subject is given in
Ref.~\cite{milc-1988-1,toussaint-1990-1,anderson-2003,johnson-2007}.

The fitting functional form suggested by the SU(2) staggered chiral
perturbation theory (SChPT) is linear as follows,
\begin{equation}
f_\text{th} (X) = \sum_{a=1}^{P} c_a F_a(X)
\label{eq:fit-func-1}
\end{equation}
where $c_a$ are the low energy constants (LECs) and
$F_a$ is a function of $X$ which represents 
collectively $X_P$ (pion squared mass of $\bar{x}x$), 
$Y_P$ (pion squared mass of $\bar{y}{y}$), and so on.
The details on $F_a$ and $X$ are given in SW-1.
Here, we focus on the X-fit of 4X3Y-NNLO fitting of the SU(2) SChPT,
which is explained in great detail in SW-1.
In this fit, we have three LECs and so $P=3$.

We are interested in the probability distribution of the average
$\bar{y}_i$ of the data $y_i(n)$.
\begin{equation}
\bar{y}_i = \frac{1}{N} \sum_{n=1}^{N} y_i(n)
\end{equation}
We assume that the measured values of $\bar{y}_i$ have a normal
distribution $P(\bar{y})$ by the central limit theorem for the
multivariate statistical analysis as follows,
\begin{equation}
P(\bar{y}) = \frac{1}{Z} \exp[ -\frac{1}{2} \sum_{i,j=1}^{D}
(\bar{y}_i - \mu_i) (N \ \Gamma^{-1}_{ij}) (\bar{y}_j - \mu_j) 
] \,,
\end{equation}
where $\mu_i$ represents the true mean value of $y_i$, which is, in
general, unknown and can be obtained as $N \rightarrow \infty$, and
\[
Z = \int [d\bar{y}] \exp[ -\frac{1}{2} \sum_{i,j=1}^{D}   
(\bar{y}_i - \mu_i) (N \ \Gamma^{-1}_{ij}) (\bar{y}_j - \mu_j) ]
\,.
\]
Here, $\Gamma_{ij}$ is the true covariance matrix, which is, in
general, unknown in our problems.
The maximum likelihood estimator of $\Gamma_{ij}$ turns out to be the
sample covariance matrix $S_{ij}$ defined as follows,
\begin{eqnarray}
S_{ij} &=& \frac{1}{N-1} \sum_{n=1}^{N} [y_i(n) - \bar{y}_i]
  [y_j(n) - \bar{y}_j]
\\
C_{ij} &=& \frac{1}{N} S_{ij},
\label{eq:cov_mat}
\end{eqnarray}
where $C_{ij}$ is the normalized sample covariance matrix.
Here, note that the covariance matrix is a symmetric and positive
definite matrix which has real and positive eigenvalues\footnote{
  Here, the positive means that the eigenvalues of the covariance
  matrix cannot be negative. In other words, some of them may be zero
  and the rest are positive.  }.
We assume that our theory\footnote{Here, it means the SU(2) SChPT.}
must describe the data well.
Then, 
\[
\mu_i \rightarrow \nu_i = f_\text{th}(X_i) = \sum_{a=1}^{P} c_a F_a(X_i)
\,.
\]
In other words, we want to test whether $\nu_i$ describes the data
reliably from the standpoint of statistics.
In this procedure, we want to determine $c_a$ (LECs) to give the best
fit.
Here, the best fit is defined by minimizing the $T^2$ of the numerical 
results $\{\bar{y}_i\}$, where the $T^2$ is defined by
\begin{eqnarray}
T^2 = \sum_{i,j=1}^{D}
[\bar{y}_i - \nu_i] [N \ S^{-1}_{ij}] [\bar{y}_j - \nu_j]
\,.
\end{eqnarray}
We notice that $Y_i = \sqrt{N} [\bar{y}_i - \nu_i]$ is distributed
according to $\mathcal{N}(\rho, \Gamma)$ and 
$\rho_i = \sqrt{N} [\mu_i - \nu_i]$.
Here, we use the same notation as in Ref.~\cite{anderson-2003}.
Then, note that $(N-1) S_{ij}$ is independently distributed as
\[ \sum_{n=1}^{N-1} Z_i(n) Z_j(n) \] where $Z(n)$ is distributed
according to $\mathcal{N}(0,\Gamma)$.
In this case, $[T^2/(N-1)] [(N-d)/d]$ is distributed as a non-central
$F$ distribution of $F_{d, N-d}$, which is defined in
Ref.~\cite{anderson-2003}, and its non-centrality parameter is
\[ 
\sum_{i,j} \rho_i \Gamma_{ij}^{-1} \rho_{j} 
=\sum_{i,j} (\mu_i - \nu_i) ( N  \ \Gamma_{ij}^{-1} ) (\mu_j - \nu_j)
\,. 
\]
Here, $d$ is the degrees of freedom of the fitting, defined by 
$d = D-P$.
In Ref.~\cite{anderson-2003}, it is proved that the limiting
distribution of $T^2$ as $N \rightarrow \infty$ is the
$\chi^2$-distribution with $P$ degrees of freedom if $\mu_i = \nu_i$.
At this point, we have to minimize $T^2$ in order to determine
the LECs: $\{c_a\}$.
Hence, we need to solve the following equation:
\begin{equation}
\frac{\partial T^2}{\partial c_a} = 0
\end{equation}
We can rewrite this equation as follows,
\begin{eqnarray}
& & \frac{\partial T^2}{\partial c_a} =
2 \sum_{i,j=1}^{D} 
\frac{\partial f_\text{th}(X_i)}{\partial c_a}
C^{-1}_{ij} [f_\text{th}(X_j) - \bar{y}_j ] = 0
\\
& & \sum_{b=1}^{P} A_{ab} \ c_b  =  h_a 
\end{eqnarray}
where
\begin{subequations}
\begin{eqnarray}
A_{ab} &=& \sum_{i,j=1}^{D} F_a (X_i) C^{-1}_{ij} F_b(X_j)
\\
h_a &=& \sum_{i,j=1}^{D} F_a (X_i) C^{-1}_{ij} \bar{y}_j \,.
\end{eqnarray}
\label{eq:mat:vec}
\end{subequations}
Here, note that the matrix $A_{ab}$ is a symmetric matrix.
The solution can be obtained by simply solving the
linear algebra.
\begin{equation}
\hat{A} \vec{c} = \vec{h} \quad \rightarrow \quad
\vec{c} = \hat{A}^{-1} \vec{h}
\label{eq:sol}
\end{equation}
So far, so good.
One caveat is that the solution of Eq.~\eqref{eq:sol} exists only if
the covariance matrix $C$ is non-singular.
In practice, even though the covariance matrix has only very small
eigenvalues, it is enough to cause a very poor fitting.
We will address this issue in the next sections.

\section{Trouble with covariance matrix for $B_K$}
\label{sec:bk-cov-mat}
%
First, we address the issue on the quality of the fitting function
form suggested by SChPT.
Second, we would like to address a typical difficulty with a general
covariance fitting of highly correlated data.
To demonstrate the problem, we choose the $B_K$ data on a coarse (C3)
ensemble out of MILC asqtad lattices using the notation of SW-1 and
SW-2.
This ensemble is particularly a good sample, because it has relatively
large statistics.
The input parameters for the C3 ensemble is summarized in 
Table \ref{tab:para-C3}.
\begin{table}[tbp]
\centering
\begin{tabular}{|l | l|}
\hline
parameter & value \\
\hline
sea quarks              & asqtad staggered fermions \\
valence quarks          & HYP staggered fermions \\
gluons                  & Symanzik improved gluon action \\
geometry                & $20^3 \times 64$ \\
number of confs         & 671 \\
number of meas          & 9 per conf \\
$am_l/am_s$             & $0.01/0.05$ \\
$1/a$                   & 1662 MeV \\
$\alpha_s$              & 0.3286 at $\mu=1/a$\\
$am_x, am_y$            & $0.005 \times n$ $(n=1,2,3,\ldots,10)$  \\
\hline
\end{tabular}
  \caption{ Parameters for the numerical study on the coarse (C3)
    ensemble. $m_l$ is the light sea quark mass, $m_s$ the strange sea
    quark mass, $m_x$ the light valence quark mass, and $m_y$ the strange
    valence quark mass. Here, ``conf'' and ``meas'' represent gauge
    configuration and measurement, respectively.}
  \label{tab:para-C3}
\end{table}

\subsection{Quality of the fitting function}
\label{subsec:qual-fit-func}
The fitting function given in Eq.~\eqref{eq:fit-func-1} is derived
from SChPT.
Hence, its reliability is directly related to the validity of SChPT.
It is beyond the scope of this paper to discuss the validity of SChPT,
which has been proved to be true through extensive numerical study in
Ref.~\cite{wlee-99,bernard-03,milc-rmp-09,steve-06,wlee-10-3}.
Here, we take a different approach to the same issue.
Basically, we take some trial fitting functions which do not have any
solid theoretical background such as SChPT but are highly empirical.
We choose two empirical fitting functions: one is linear and the other
is quadratic in 
\[
X = \frac{m_\pi^2(\bar{x}x;\xi_5)}{\Lambda^2} \,,
\] 
where $\Lambda = 1.0$GeV (a generic scale of chiral perturbation theory).
The fitting results are summarized in Table \ref{tab:fit-func}.
\begin{table}[tbp]
\centering
\begin{tabular}{| c | c | r r |}
\hline
$f(X)$ & memo & $\chi^2_\text{diag}$ & $\chi^2$ \\
\hline
$c_1 + c_2 X$           & trial & 1.47(73) & 12.6(50)  \\
$c_1 + c_2 X + c_3 X^2$ & trial & 0.19(13) &  8.5(59)  \\
$f_\text{th}(X)$        & SChPT & 0.16(12) &  7.2(54)  \\
\hline
\end{tabular}
  \caption{ List of fitting functions and its quality.
    $f(X)$ is the fitting function. The ''memo'' means 
    the theoretical origin of the fitting functions.
    $\chi^2_\text{diag}$ represents $T^2$ with diagonal approximation
    and $\chi^2$ represents $T^2$ with the full covariance matrix.}
  \label{tab:fit-func}
\end{table}
From the table, we observe that the $\chi^2$ value for $f_\text{th}$
is consistently smallest, which indicates that our choice of the
fitting function is quite optimal.

We will get back to this issue when we discuss about the full
covariance fitting and its trouble.

\subsection{Full covariance fitting}
\label{subsec:full-cov-fit}
As one can see in Table \ref{tab:para-C3}, we have 55 combinations of 
$m_x$ and $m_y$ (10 degenerate pairs ($m_x=m_y$) and 45 non-degenerate
pairs ($m_x \ne m_y$)), and so $D=55$ for this example.
Since the number of gauge configurations is 671, $N=671$.
The fit type is the X-fit of the 4X3Y-NNLO fitting of the SU(2) SChPT
as explained in SW-1, and so $P=3$.
In a single X-fit, we use only 4 data points and so we may say that
$D=4$.
Now, let us consider the $D \times D$ covariance matrix.
It has only 10 ($=D(D+1)/2$) components which can be determined
completely in a linearly independent way, using 671 independent
configurations.
Here, we focus on the troublesome small eigenvalues of the covariance
matrix.

To be concrete, let us walk through a specific example of $B_K$ to
demonstrate the problem and its consequence.
In the X-fit, we fix $am_y = 0.05$ and select 4 data points
of $am_x =$ 0.005, 0.010, 0.015, 0.020 to fit to the functional
form suggested by the SU(2) SChPT as in SW-1.
Hence, the covariance matrix $C_{ij}$ is a $4 \times 4$ matrix.
\begin{equation}
C_{ij} = \left[ \begin{array}{c c c c}
   1.42,   & 0.661, & 0.398, & 0.274 \\
   0.661,  & 0.392, & 0.271, & 0.204 \\
   0.398,  & 0.271, & 0.205, & 0.165 \\
   0.274,  & 0.204, & 0.165, & 0.138
\end{array} \right] \times 10^{-5}
\end{equation}
Its eigenvalues are
\begin{eqnarray}
\lambda_i = \{\ 1.95\times 10^{-5}, \  1.92\times 10^{-6}, 
  \  7.58\times 10^{-8},\  1.11\times 10^{-9}
\}
\,.
\end{eqnarray}
The components of the matrix $C_{ij}$ are 
between $1.42 \times 10^{-5} $ and $1.38 \times 10^{-6}$.
In the meanwhile, the smallest eigenvalue is smaller than the
components by three orders of magnitude.

Now let us look into the eigenvectors:
\begin{eqnarray}
v_1 = \left[ \begin{array}{r}
   0.837 \\
   0.429 \\
   0.276 \\
   0.200
\end{array} \right]\,, \quad
v_2 = \left[ \begin{array}{r}
   -0.508 \\
    0.387 \\
    0.542 \\
    0.546
\end{array} \right]\,, \quad 
v_3 = \left[ \begin{array}{r}
    0.202 \\
   -0.739 \\
    0.0725 \\
    0.639
\end{array} \right]\,, \quad
v_4 = \left[ \begin{array}{r}
    -0.0378 \\
     0.347 \\
    -0.790 \\
     0.503
\end{array} \right]\,.
\end{eqnarray}
The eigenvector $v_4$ corresponds to the smallest eigenvalue.
This eigenmode dominates the $\chi^2$ fitting completely.
Let us expand $\bar{y}$ in terms of eigenvectors as follows,
\begin{equation}
\bar{y} = \sum_{i=1}^4 a_i v_i \,,
\end{equation}
where $a_i$ is the eigenmode projection coefficient.
\begin{table}[tbp]
\centering
\begin{tabular}{| c | c c c c |}
\hline
$i$ & 1 & 2 & 3 & 4 \\
\hline
$a_i$      & 1.021(4)   & 0.5655(14) & 0.1061(3)    & 0.01442(3)    \\
$\alpha_i$ & 0.7589(18) & 0.2328(18) & 0.008190(69) & 0.0001513(12) \\
\hline
\end{tabular}
  \caption{ Eigenmode projection coefficients for $\bar{y}$.}
  \label{tab:y-eigen}
\end{table}
Let us define $\alpha_i$ as follows,
\[
\alpha_i = \frac{|a_i|^2}{\sum_j |a_j|^2 } \,.
\]
Here, note that $\alpha_i$ represents the probability of the specific
eigenmode in the data $\bar{y}$.
In Table \ref{tab:y-eigen}, we show $a_i$ and $\alpha_i$
for the data $\bar{y}$.
The eigenmode $v_1$ and $v_2$ describes more than 99\% of 
the data $\bar{y}$.
The remaining 0.8\% is occupied by $v_3$ and the rest 0.015\%
is from $v_4$.

We can rewrite the inverse covariance matrix as follows,
\begin{equation}
[C_{ij}^{-1}] = \sum_{k=1}^4 \frac{1}{\lambda_k} 
| v_k \rangle \langle v_k |
\,.
\end{equation}
Hence, we can also express the $T^2$ as follows,
\begin{equation}
T^2 = \sum_{i=1}^4 \frac{1}{\lambda_i}  \langle \Delta y | v_i \rangle^2
\,,
\end{equation}
where $\Delta y_j \equiv [\bar{y}_j - \nu_j]$.
Here, note that the least $T^2$ fitting is completely dominated by
$v_4$, which becomes an inconvenient truth and causes a serious
trouble in covariance fitting.
\begin{figure}[tbp]
\centering
  \includegraphics[width=20pc]
  {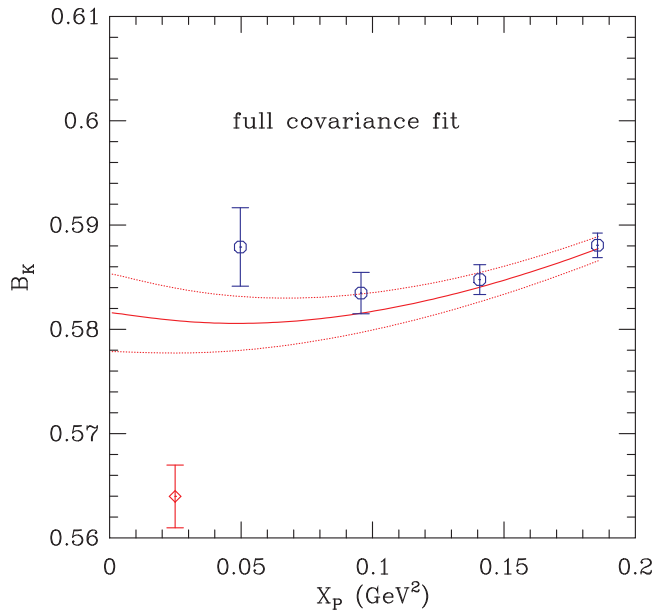}
  \caption{ $B_K(1/a)$ vs. $X_P$ on the C3 ensemble.  The fit type is
    4X3Y-NNLO in the SU(2) analysis.  We fix $am_y = 0.05$.  The red
    line represents the results of fitting with the full covariance
    matrix.  The red diamond corresponds to the $B_K$ value
    obtained by extrapolating $m_x$ to the physical light valence
    quark mass after setting all the pion multiplet splittings to
    zero.}
  \label{fig:cov-C3}
\end{figure}
In Figure \ref{fig:cov-C3}, we show the fitting results with the 
full covariance matrix.
As one can see, the fitting curve does not pass through the data
points.
Hence, the quality of fitting looks poor to our eyes.
The $T^2$ value is
\[
T^2 = 7.2 \pm 5.4
\,.
\]
The multivariate statistical theory predicts the following 
\cite{schervish-1995}:
\begin{eqnarray}
{\cal E} (T^2) &=& (d + \kappa) \left[ 1 + \frac{d+1}{N} + 
  {\cal O}(\frac{1}{N^2}) \right] \\
{\cal V} (T^2) &=& 2 (d + 2\kappa) 
  \biggl[ 1 + \frac{1}{N} 
    \Big(2 d + 4 + \frac{(d+\kappa)^2}{d + 2 \kappa} \Big)
    + {\cal O}(\frac{1}{N^2} ) \biggl] 
\end{eqnarray}
where ${\cal E} (T^2)$ and ${\cal V} (T^2)$ represent the expectation 
value and variance of the $T^2$ distribution.
Here, $d$ is the degrees of freedom and $\kappa$ is the non-centrality
parameter.
If the degrees of freedom is comparable to the number of samples ($d
\approx N$), the leading deviation of the $T^2$ distribution from the
$\chi^2$ distribution becomes of order ${\cal O}(1)$ and so we can not
use the $\chi^2$ distribution in this case, which is also pointed out
in Ref.~\cite{michael-1994-1} in the context of distorted normal
distribution.
In Ref.~\cite{toussaint-08}, the sample size effect is systematically
explained in terms of $1/N$ expansion.

However, in our example of $B_K$, $d=1$ and $N=671$.
Hence, the leading correction of the $T^2$ distribution to the
$\chi^2$ distribution will be negligibly small ($\approx 0.15$\%).
In our example, the non-centrality parameter can be 
estimated by $\kappa \approx T^2 - d = 6.2$.
Then, we can obtain ${\cal V}(T^2)$ as follows,
\begin{eqnarray}
{\cal V}(T^2) &\approx& 2 (d + 2\kappa)
= 26.8 
\end{eqnarray}
Hence, the error of $T^2$ is supposed to be $\sqrt{26.8} = 5.2$,
which is reasonably consistent with the measured value 5.4.
The $\kappa$ is the non-centrality parameter which represents
how much the fitting function deviates from the true mean values.
In our example, $\kappa = 6.2$ turns out to be a rather large value
which comes from the fact that the small deviation of our fitting
function from the true value due to the truncation of the higher order
terms in the series expansion of the SU(2) SChPT can be amplified
dramatically if there are small eigenvalues in the covariance matrix.

Let us decompose the fitting function in terms of eigenmodes
as follows,
\begin{eqnarray}
\nu &=& f_\text{th} = \sum_{i=1}^4 b_i v_i
\\
\beta_i &=& \frac{|b_i|^2}{\sum_j |b_j|^2 } \,.
\end{eqnarray}
To see how much the fitting function deviates from the data in 
a specific eigenmode, we define the difference, $\delta_i$,
as follow,
\begin{equation}
  \delta_i = |a_i - b_i|.
\label{eq:delta}
\end{equation}
As summarized in Table \ref{tab:f-eigen-cov}, the difference, 
$\delta_i$, is $7.22\times10^{-3}$ for $v_1$, $2.40\times10^{-3}$ for 
$v_2$, $3.28\times10^{-4}$ for $v_3$, whereas it is only 
$9.69\times10^{-6}$ for $v_4$.   
Hence, the procedure of the least $\chi^2$ fitting works hard for the
coefficient of $v_4$ but work less precisely for the coefficients of
$v_1$ and $v_2$ mainly because the eigenvalue $\lambda_4$ is
significantly smaller than $\lambda_1$ and $\lambda_2$.
The irony is that the data has only 0.015\% overlap with $v_4$ while
more than 99\% of it is dominated by $v_1$ and $v_2$.
In this sense, the failure of the full covariance fitting is obviously
due to the fact that the least $\chi^2$ fitting tries to determine the
coefficient of $v_4$ component of the data very precisely but lose 
precision in determining the coefficients of the $v_1$ and $v_2$ 
components.
As a consequence, the fitting curve misses the data points and
the quality of fitting looks poor to our eyes. 
\begin{table}[tbp]
\centering
\begin{tabular}{| c | c c c c |}
\hline
$i$ & 1 & 2 & 3 & 4 \\
\hline
$b_i$                 & 1.014(4)   & 0.5679(11) & 0.1058(3)    & 0.01443(3)    \\
$a_i$                 & 1.021(4)   & 0.5655(14) & 0.1061(3)    & 0.01442(3)    \\
$10^5 \cdot \delta_i$ & 722(270)   & 240(90)    & 32.8(123)    & 0.969(362)    \\
$\beta_i$             & 0.7548(10) & 0.2368(9)  & 0.008212(69) & 0.0001528(11) \\
\hline
\end{tabular}
  \caption{ Eigenmode decomposition of $f_\text{th}$ for the
    full covariance fitting.}
  \label{tab:f-eigen-cov}
\end{table}

How can we improve the situation?
We will address this issue in the next section.

\section{Prescription}
\label{sec:pres}
Here, we present a number of potential solutions to
the problem raised in the previous section.
Part of the solutions are well-known but vulnerable.
Part of the solutions are new but of noteworthy merit.
We will go through them one by one, and discuss about the
pros and cons.

\subsection{Diagonal approximation}
\label{subsec:diag}
\begin{figure}[tbp]
\centering
  \includegraphics[width=20pc]
  {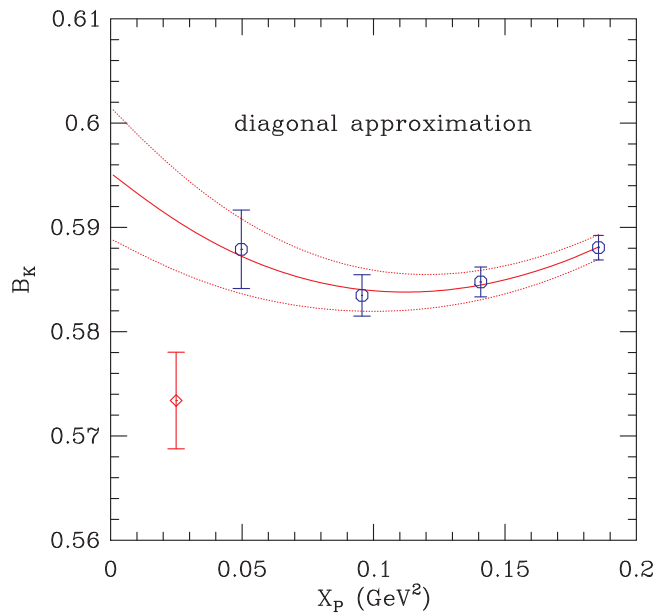}
  \caption{ $B_K(1/a)$ vs. $X_P$ on the C3 ensemble.  All the
    parameters are the same as in
    Figure \protect\ref{fig:cov-C3}. Here, the red line represents the
    results of the uncorrelated fitting using the diagonal
    approximation. }
  \label{fig:diag-C3}
\end{figure}

One simple solution to the problem is to use the diagonal
approximation (uncorrelated fitting) \cite{michael-1994-1}.
In this method, we neglect the off-diagonal covariance as follows:
\begin{eqnarray}
C_{ij} = 0 \qquad \text{ if } i \ne j \,.
\end{eqnarray}
In this way, the small eigenvalue problem disappears and the fitting
results are shown in Figure \ref{fig:diag-C3}.

The fitting function $f_\text{th}$ in the diagonal approximation is
decomposed into eigenmodes of the full covariance matrix in Table
\ref{tab:f-eigen-diag}.
In this fit, the difference, $\delta_i$, is $3.80\times10^{-4}$ for
$v_1$, $4.26\times10^{-4}$ for $v_2$, $5.23\times10^{-4}$ for $v_3$
and $4.85\times10^{-4}$ for $v_4$.
%
%
Here, note that the diagonal approximation method removes the small
eigenvalues and so it takes all the eigenmodes, equally.
As a result, the differences for all directions are less than or equal 
to $5.23\times10^{-4}$.
Hence, the fitting looks quite reasonable to our eyes as one can see
in Figure \ref{fig:diag-C3}.
\begin{table}[tbp]
\centering
\begin{tabular}{| c | c c c c |}
\hline
$i$ & 1 & 2 & 3 & 4 \\
\hline
$b_i$                 & 1.021(4)   & 0.5659(13) & 0.1056(3)    & 0.01490(18)   \\
$a_i$                 & 1.021(4)   & 0.5655(14) & 0.1061(3)    & 0.01442(3)    \\
$10^5 \cdot \delta_i$ & 38.0(142)  & 42.6(159)  & 52.3(195)    & 48.5(181)     \\
$\beta_i$             & 0.7586(17) & 0.2332(16) & 0.008112(77) & 0.0001617(35) \\
\hline
\end{tabular}
  \caption{ Eigenmode decomposition of $f_\text{th}$ for the
    fitting with diagonal approximation.}
  \label{tab:f-eigen-diag}
\end{table}

In the diagonal approximation, we lose the physical meaning of
$\chi^2$, because we underestimate the $\chi^2$ by neglecting
the off-diagonal terms in the covariance matrix.
This point is a major drawback of the diagonal approximation.

\subsection{Cutoff method}
\label{subsec:cutoff}
\begin{figure}[tbp]
\centering
  \includegraphics[width=20pc]
  {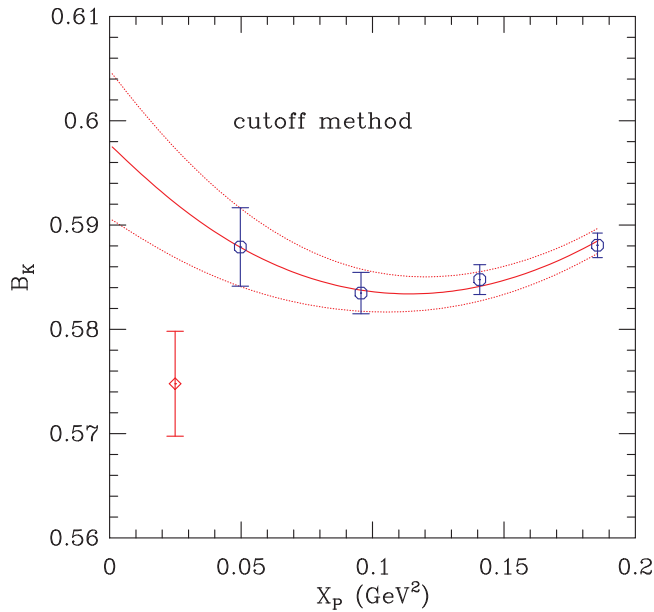}
  \caption{ $B_K(1/a)$ vs. $X_P$ on the C3 ensemble.  All the
    parameters are the same as in
    Figure \protect\ref{fig:diag-C3}. Here, the red line represents the
    results of the covariance fitting after removing the smallest
    eigenvalue using the cutoff method. }
  \label{fig:cutoff-C3}
\end{figure}

Another solution is to exclude the $v_4$ eigenmode from the fitting.
Since we know that the $v_4$ eigenmode has least significant
contribution in the data, we can think of the philosophy of removing
them by hand as suggested by Refs.~\cite{NR-2007-1,kilcup-1994-1}.
A popular and systematic way of chopping away the $v_4$ eigenmode is
to set up such a cutoff that we project out the eigenvectors of those
eigenvalues smaller than the cutoff into a null space of the inverse
covariance matrix $C^{-1}_{ij}$.
We call this the cutoff method.
A number of lattice QCD groups
\cite{ref:fnal-2010-1,ref:fnal-2010-2,ref:lanl-1999-1} use this method
in the popular name of the SVD (singular value decomposition) method.
Now let us walk through an example to demonstrate how it works.
In our example of $B_K$, we have three parameters to determine
from the fit. 
Hence, it is possible to remove only one eigenmode $v_4$ out of the
four by setting $1/\lambda_4 = 0$.
In Figure \ref{fig:cutoff-C3}, we show the results of the covariance
fitting using the cutoff method.
It is amusing to see how good it works.
The results are consistent with those in Figure \ref{fig:diag-C3}.
\begin{table}[tbp]
\centering
\begin{tabular}{| c | c c c c |}
\hline
$i$ & 1 & 2 & 3 & 4 \\
\hline
$b_i$                 & 1.021(4)   & 0.5655(14) & 0.1061(3)    & 0.01524(30)   \\
$a_i$                 & 1.021(4)   & 0.5655(14) & 0.1061(3)    & 0.01442(3)    \\
$10^5 \cdot \delta_i$ & 0(0)       & 0(0)       & 0(0)         & 82.1(307)     \\
$\beta_i$             & 0.7589(18) & 0.2328(18) & 0.008190(69) & 0.0001690(63) \\
\hline
\end{tabular}
  \caption{ Eigenmode decomposition of $f_\text{th}$ for the
    fitting with the cutoff method.}
  \label{tab:f-eigen-cutoff}
\end{table}
In Table \ref{tab:f-eigen-cutoff} we decompose the fitting function
$f_\text{th}$ obtained using the cutoff method into eigenmodes of
the full covariance matrix.
In this case, the difference, $\delta_i$, is zero for $i=1,2,3$ and
is $8.21\times 10^{-4}$ in $v_4$.
In this method, we do not care about the $v_4$ eigenmode at all.
As a result, the fitting quality looks quite good to our eyes, which
is quite consistent with that of the diagonal approximation.

However, a major drawback of the cutoff method is that we cannot give
the physical meaning to the quality of fit, which is normally
reflected in the minimized value of $\chi^2$.
In Appendix~\ref{app:subsec:dist_chisq_cutoff}, we show that the
probability distribution of $\chi^2$ defined in cutoff method is the
$\chi^2$ distribution with degrees of freedom equal to $D-P-R$.
Here, $D$ is the number of data points, $P$ is the number of fitting
parameters and $R$ is the number of removed eigenmodes.
However, even though we know the distribution of the $\chi^2$ in
cutoff method, we cannot measure the quality of fit through the 
minimized value of $\chi^2$.
This is because we remove some eigenmodes and the
$\chi^2$ in the cutoff method is orthogonal to the removed eigenmodes.
As you know, the physical $\chi^2$ has $D-P$ degrees of freedom, while
the $\chi^2$ of the cut-off method possesses $D-P-R$ degrees of
freedom.
Unfortunately, the missing degrees of freedom, $R$ are physical.

The situation becomes even worse when we use the resampling method,
such as jackknife method or bootstrap method.
When the size of covariance matrix is large, we might lose a
control over the number of the small eigenvalues that we remove.
In other words, in one jackknife sample, we may remove two small
eigenvalues and in another jackknife sample, we may remove three of
them.
During the procedure, the definition of $\chi^2$ is shifting from 
one to another.

\subsection{Modified covariance matrix}
\label{subsec:mod-cov-mat}

One may take another approach to handle the small eigenmodes
as in Ref.~\cite{milc-2002-1}.
We first define the correlation matrix $\widetilde{C}$ as follows,
\begin{eqnarray}
\sigma_i &=& \sqrt{C_{ii}}
\\
\widetilde{C}_{ij} &=& \frac{C_{ij}}{\sigma_i \sigma_j}
\end{eqnarray}
such that the diagonal components of $\widetilde{C}_{ij}$ is 1.
In our example of $B_K$, it is
\begin{equation}
\widetilde{C}_{ij} = \left[ \begin{array} {c c c c}
   1.000 & 0.888 &  0.738 &  0.619 \\
   0.888 & 1.000 &  0.955 &  0.877 \\
   0.738 & 0.955 &  1.000 &  0.978 \\
   0.619 & 0.877 &  0.978 &  1.000
   \end{array}
\right]
\end{equation}
Hence, we can say that the correlation matrix is a normalized
covariance matrix.
In our example, the off-diagonal terms are quite large between
0.6 and 1.0, which indicates that the data is highly correlated
and moving together in the same direction.
The eigenvalues of the correlation matrix are
\begin{eqnarray}
\widetilde{\lambda}_i &=& \{
 \  3.54, \  0.437, \  0.0249, \  0.000521 \}
\end{eqnarray}
One may choose their cutoff as $\lambda_\text{cut} = 1$
as in Ref.~\cite{milc-2002-1}.
We remove by hand all the eigenmodes whose eigenvalue is 
smaller than $\lambda_\text{cut}$.
In our example, this corresponds to removing three eigenmodes of
$\widetilde{\lambda}_2$, $\widetilde{\lambda}_3$, and
$\widetilde{\lambda}_4$.
It is obvious that the remaining correlation matrix is highly
singular.
In order to avoid the singular behavior, we restore the diagonal
components back to 1 by hand as in Ref.~\cite{milc-2002-1}.
Let us call this modified correlation matrix $\overline{C}_{ij}$.
Then, let us define the modified covariance matrix $M_{ij}$ as
\begin{equation}
M_{ij} = \overline{C}_{ij} \times \sigma_i \times \sigma_j
\end{equation}
In our example of $B_K$, it is
\begin{equation}
M_{ij} = \left[ \begin{array} {c c c c}
   1.42  &  0.632 &  0.453 &  0.353 \\
   0.632 &  0.392 &  0.275 &  0.214 \\
   0.453 &  0.275 &  0.205 &  0.153 \\
   0.353 &  0.214 &  0.153 &  0.138
   \end{array}
\right] \times 10^{-5}
\end{equation}
Then, we use $M_{ij}$ as their covariance matrix and
fit the data.
We call this the modified covariance matrix (MCM) method.\footnote{ MILC
  does not use this method anymore in their fitting
  \cite{milc-private-2011}.}

\begin{figure}[tbp]
\centering
  \includegraphics[width=20pc]
  {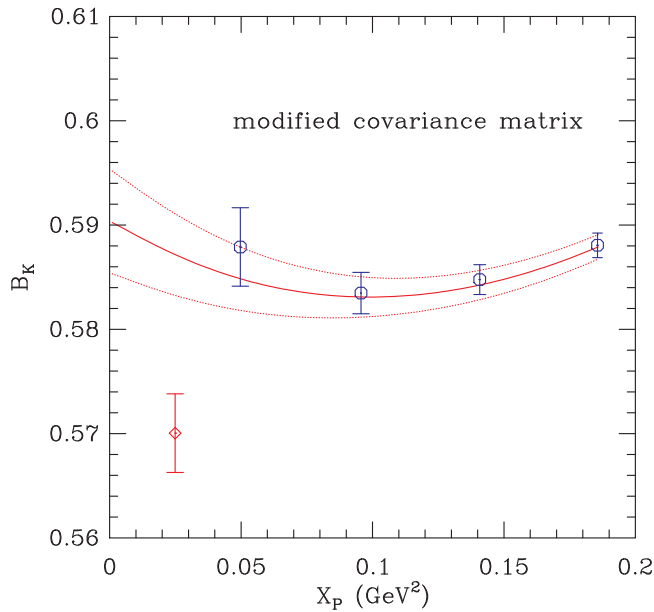}
  \caption{ $B_K(1/a)$ vs. $X_P$ on the C3 ensemble.  All the
    parameters are the same as in
    Figure \protect\ref{fig:diag-C3}. Here, the red line represents the
    results of the covariance fitting using the modified covariance
    matrix. }
  \label{fig:mod-C3}
\end{figure}

In Figure \ref{fig:mod-C3}, we show the results of the
covariance fitting using the modified covariance matrix.
The fitting quality is somewhere between the diagonal
approximation and the full covariance fitting.
\begin{table}[tbp]
\centering
\begin{tabular}{| c | c c c c |}
\hline
$i$ & 1 & 2 & 3 & 4 \\
\hline
$b_i$                 & 1.018(4)   & 0.5665(12) & 0.1056(3)    & 0.01473(12)   \\
$a_i$                 & 1.021(4)   & 0.5655(14) & 0.1061(3)    & 0.01442(3)    \\
$10^5 \cdot \delta_i$ & 289(108)   & 103(38)    & 48.9(182)    & 30.9(116)     \\
$\beta_i$             & 0.7573(13) & 0.2344(13) & 0.008143(73) & 0.0001584(24) \\
\hline
\end{tabular}
  \caption{ Eigenmode decomposition of $f_\text{th}$ for the
    fitting with the MCM method.}
  \label{tab:f-eigen-mod}
\end{table}
In Table \ref{tab:f-eigen-mod} we decompose the fitting function
$f_\text{th}$ obtained using the MCM method into eigenmodes of
the full covariance matrix.
%
%
In this case, the difference, $\delta_i$, is $2.89\times10^{-3}$ for 
$v_1$, $1.03\times10^{-3}$ for $v_2$, $4.89\times10^{-4}$ for $v_3$
and $3.09\times10^{-4}$ for $v_4$.
As a result, the fitting quality look mediocre to our eyes and worse
than that of the diagonal approximation.

The main disadvantage of using the MCM method is that the $\chi^2$
loses the physical meaning completely, because the modified covariance
matrix is significantly different from the full covariance matrix by
construction.
However, one may view the diagonal approximation as a special case of
the MCM method where all the eigenmodes are removed.
In this sense, the MCM method is as bad as the diagonal approximation.

\subsection{Eigenmode shift method}
\label{subsec:shift}

So far, all the methods are based on the philosophy that we may
manipulate or modify the covariance matrix.
This philosophy is very dangerous in a sense that the modification
results in a missing information in physics.
The missing information is highly physical.
Hence, it is not a good idea to modify the covariance matrix.
The only degrees of freedom that we have is to modify the fitting
function, but not the covariance matrix.

We know that the whole trouble comes from the inexact fitting
function that has small error in $v_4$ eigenmode direction.
Hence, we can think of a new fitting function $f_\text{th}'$
defined as follows,
\begin{equation}
f_\text{th}'(X) = f_\text{th} (X) + \eta v_4
\end{equation}
Here, $\eta$ is a tiny parameter which can be determined
using the Bayesian method.
The Bayes theorem \cite{sivia-2006} says that
\begin{eqnarray}
P(A|\mathbb{D},I) &\propto& P(\mathbb{D} | A,I) \times P(A|I)
\end{eqnarray}
Here, $A$ represents our theoretical hypothesis, $\mathbb{D}$ is the
data, and $I$ corresponds to the background information.
Note that $P(\mathbb{D} | A,I)$ means the probability that we obtain
the data set of $\mathbb{D}$ if $A$ and $I$ are given to us.
We know the conditional likelihood function of $P(\mathbb{D} | A,I)$,
which is nothing but
\begin{eqnarray}
P(\mathbb{D} | A,I) &\propto& \exp\left(-\frac{\chi^2}{2}\right) 
\\
\label{eq:chisq_ES}
\chi^2 &=& \sum_{i,j} [\bar{y}_i - \nu_i'] C^{-1}_{ij} [\bar{y}_j - \nu_j']
\\
\nu_i' &=& f_\text{th}'(X_i)\,,
\end{eqnarray}
as explained in Ref.~\cite{sivia-2006}.

In addition, if we impose the maximum entropy principle
\cite{sivia-2006} on the prior condition that $ a_\eta - \sigma_\eta
\lesssim \eta \lesssim a_\eta + \sigma_\eta$, then the prior becomes
the following:
\begin{eqnarray}
P(A | I) &\propto& \exp\left( - \frac{\chi^2_\textrm{prior}}{2} \right)
\\
\chi^2_\textrm{prior} &=& \frac{(\eta-a_\eta)^2}{\sigma_\eta^2}
\end{eqnarray}
Then, we obtain the posterior pdf as follows:
\begin{eqnarray}
P(A|\mathbb{D},I) &\propto& 
\exp\left( - \frac{\chi^2_\textrm{aug}}{2}\right)
\\
\label{eq:aug_chisq_ES}
\chi^2_\text{aug} &=& \chi^2 + \chi^2_\textrm{prior}
\end{eqnarray}
The Bayesian principle \cite{sivia-2006} is to determine the fitting
parameters by maximizing the posterior pdf: $P(A|\mathbb{D},I)$.
This is equivalent to minimizing the $\chi^2_\textrm{aug}$.

Let us switch the gear to our choice of the prior condition.
From the SChPT, the neglected highest order term in the $f_\text{th}
(X)$ is
\[
X^2 (\ln(X))^2 \approx 0.006
\]
where $X = X_P/\Lambda^2 \approx 0.02$.
The only constraint on the coefficient $c_4$ of this term is that
$c_4 = 0 \pm 1$, but we do not know the sign of $c_4$.
Hence, we set $a_\eta = 0$ and $\sigma_\eta = 0.006$.

Then we perform the full covariance fitting with the extra fitting
parameter, $\eta$, by minimizing $\chi^2_\text{aug}$ (the Bayesian
principle).
The obtained $\eta$ is
\[
\eta = -0.00082(31)
\,.
\]
When we do the extrapolation to the physical pion mass, we use only 
the $f_\text{th}(X)$ function, dropping out the $\eta$ term, which 
is too small to make any difference at any rate.
We call this the eigenmode shift (ES) method.
\begin{figure}[tbp]
\centering
  \includegraphics[width=20pc]
  {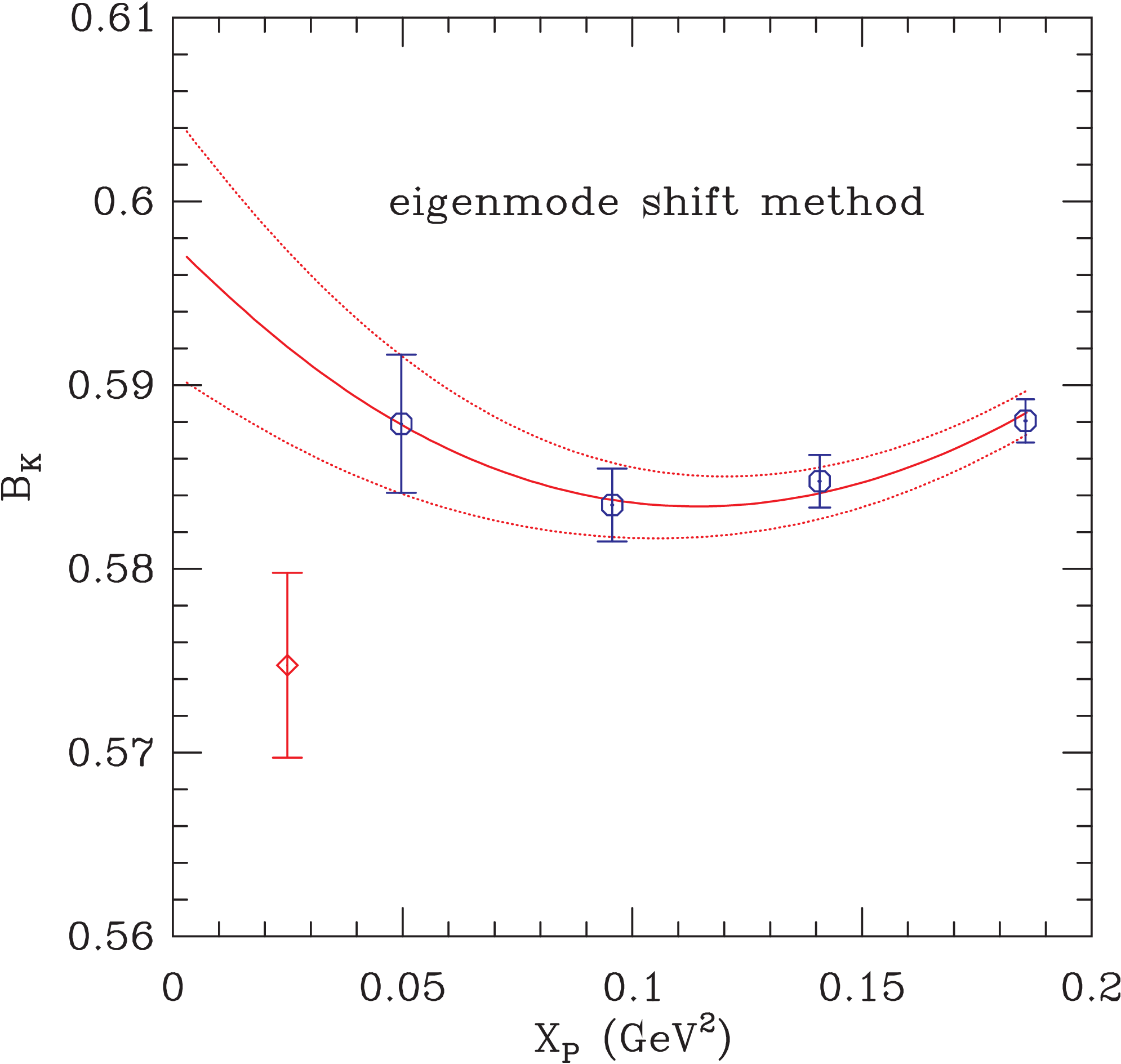}
  \caption{ $B_K(1/a)$ vs. $X_P$ on the C3 ensemble.  All the
    parameters are the same as in
    Figure \protect\ref{fig:diag-C3}. Here, the red line represents the
    results of the eigenmode shift method.}
  \label{fig:shift-C3}
\end{figure}
In Figure \ref{fig:shift-C3}, we show the fitting results obtained 
using the ES method.
In our example, tiny correction proportional to $\eta$ makes the
fitting results that pass through the average data point.
\begin{table}[tbp]
\centering
\begin{tabular}{| c | c c c c |}
\hline
$i$ & 1 & 2 & 3 & 4 \\
\hline
$b_i$                 & 1.021(4)   & 0.5655(14)  & 0.1061(3)    &
0.01524(30)   \\
$a_i$                 & 1.021(4)   & 0.5655(14)  & 0.1061(3)    &
0.01442(3)    \\
$10^5 \cdot \delta_i$ & 1.88(70)   & 0.624(233)  & 0.0855(319)  & 81.9(306)   \\
$\beta_i$             & 0.7589(18) & 0.2328(18)  & 0.008190(69) &
0.0001690(63) \\
\hline
\end{tabular}
 \caption{ Eigenmode decomposition of $f_\text{th}$ for the
   fitting with the ES method.}
 \label{tab:f-es}
\end{table}
%
%
In Table \ref{tab:f-es}, we show the eigenmode decomposition of 
$f_\textrm{th}$ when we use the ES method in fitting.
As one can see in the table, the $\delta_i$ is smaller by order of magnitude
compared with the diagonal approximation for $i=1,2,3$.
The only non-trivial component is $\delta_4$, which is taken care of by
the shift parameter $\eta$.
As a result, in Table \ref{tab:f'-es}, the $\delta_4$ for
$f_\textrm{th}'$ becomes negligibly small by the shift parameter
$\eta$.
\begin{table}[tbp]
\centering
\begin{tabular}{| c | c c c c |}
\hline
$i$ & 1 & 2 & 3 & 4 \\
\hline
$b_i$                 & 1.021(4)   & 0.5655(14)  & 0.1061(3)    &
0.01442(3)   \\
$a_i$                 & 1.021(4)   & 0.5655(14)  & 0.1061(3)    &
0.01442(3)    \\
$10^5 \cdot \delta_i$ & 1.88(70)   & 0.624(233)  & 0.0855(319)  &
0.00252(94)   \\
$\beta_i$             & 0.7589(18) & 0.2328(18)  & 0.008190(69) &
0.0001513(12) \\
\hline
\end{tabular}
 \caption{ Eigenmode decomposition of $f_\text{th}'$ for the
   fitting with the ES method.}
 \label{tab:f'-es}
\end{table}
The success of the fitting can be checked against the hypothesis
that $|\eta| \lesssim 0.006$.
The results of fitting say that $\eta = -0.00082(31)$, which is highly
consistent with the hypothesis.
In this sense, the Bayesian prior is quite reasonable and self-consistent.

Let us turn to the issue of quality of fitting and physical
interpretation of $\chi_\textrm{aug}^2$.
In a naive sense of physical interpretation, we may count the prior
condition as one of data points, and we consider the shift parameter
$\eta$ as a new parameter in the fitting.
Hence, in this interpretation, the effective number of data points
is $\tilde{D} = D + 1$, and the effective number of unknown parameters
of the fitting function is $\tilde{P} = P + 1$.
Accordingly, the effective degrees of freedom becomes $\tilde{d} = 
\tilde{D} - \tilde{P} = D - P = d$.

Let us redefine the notation for the eigenmodes as follows:
\begin{eqnarray}
\tilde{v}_i &=& \begin{bmatrix}
  v_i \\
  0
  \end{bmatrix} 
\quad \text{for} \quad i= \{1,2,3,4\}
\\
\tilde{v}_5 &=& \begin{bmatrix}
  0 \\
  0 \\
  0 \\
  0 \\
  1
  \end{bmatrix} 
\end{eqnarray}
and
\begin{eqnarray}
\tilde{\lambda}_i &=& \lambda_i \quad \text{for} \quad i= \{1,2,3,4\}
\\
\tilde{\lambda}_5 &=& \sigma_\eta^2
\end{eqnarray}
Then, we can rewrite the $\chi^2_\textrm{aug}$ as follows:
\begin{eqnarray}
\chi^2_\textrm{aug} = \sum_{a=1}^5 \frac{\delta_a^2}{\tilde{\lambda}_a}
\label{eq:chisq-dof}
\end{eqnarray}
where $\delta_a$ is defined similarly as in Eq.~\eqref{eq:delta}.
We can prove with ease that all the eigenvectors are orthogonal
to each other.
If we assume that we may count the prior condition as one of the
data points, we realize that the effective degrees of freedom have been
increased by one as we observe in Eq.~\eqref{eq:chisq-dof}.
The number of unknown parameters in fitting is $\tilde{P}$.
Hence, even in the strict sense of physical interpretation, the
effective degrees of freedom is $\tilde{d} = \tilde{D}-\tilde{P}$.
Therefore, the $\chi^2_\textrm{aug}$ must follow the normal $\chi^2$
distribution with $\chi^2_\textrm{aug} \approx \tilde{d} \pm \sqrt{2
  \tilde{d}} = d \pm \sqrt{2 d}$, as explained in Appendix
\ref{app:subsec:dist_chisq_ES}.
%

However, in our Bayesian prior information, we did not use the
statistical information for $a_\eta$ and $\sigma_\eta$, but the 
optimal range of possible value of $\eta$ to set the $a_\eta$ and 
$\sigma_\eta$. ({i.e.} $a_\eta \ne \mathcal{E}(\eta)$ and 
$\sigma_\eta \ne \sqrt{\mathcal{V}(\eta)}$).
Hence, the choice of $a_\eta$ and $\sigma_\eta$ could be larger
or smaller (overestimated or underestimated) than the statistical 
value of them.
As a consequence, our estimate of $\chi^2_\textrm{aug}$
could be smaller or larger than the normal value of 
$\chi^2_\textrm{aug} \approx \tilde{d}$.
Hence, in this approach of Bayesian method, we cannot rely on the
strict statistical interpretation of $\chi^2_\textrm{aug}$.
One good news is that the probability interpretation is still possible
with $\chi^2_\textrm{aug}$, which allows for the quality of
fitting\footnote{Here, the quality of fitting means that we can
  compare two different fitting procedures and determine which fitting
  is more reliable based on the Bayesian method. For example, we can
  compare the full covariance fitting and the ES method for $B_K$,
  since both of these methods allows for the probability
  interpretation.} and the model selection on the basis of the
Bayesian statistics \cite{sivia-2006}.
Hence, we can, at least, tell from $\chi^2_\textrm{aug}$ which model
or hypothesis is more probable.
This is an important point because it justifies the fitting procedure
that finds the most probable fitting parameters by minimizing
$\chi^2_\textrm{aug}$.

In our example of $B_K$, the $\chi^2_\text{aug} / \text{dof}$ for
the ES method is $0.019(14)$.

In the limit of $\sigma_\eta \rightarrow \infty$, we can
remove the prior condition completely, which we call 
unconstrained ES method.
In Appendix~\ref{app:sec:equiv_cutoff_ES}, we prove that
the unconstrained ES method is equivalent to the cutoff method.
However, the original ES method is quite different from the
cutoff method in the following sense.
The effective number of degrees of freedom for the ES method is 
$\tilde{d} = \tilde{D} - \tilde{P} = 1$ while that for the cutoff
method is $\tilde{d} = (D-1) - P = 0$.
In addition, the ES method is rigorously based on the Bayesian
method and is subject to the probability interpretation.
However, the cutoff method does not allow for the probability
interpretation mainly because it modifies the Hilbert space for
the covariance matrix.
In addition, in the ES method, we modify the fitting function
by the shift parameter $\eta$, which we can monitor and gives
us an estimate of how much we are changing the fitting function.
In the case of the cutoff method, we do not know how much of the
fitting function we are dumping into the null space of the covariance
matrix.
In order to illustrate the difference between the ES method and the
cutoff method, we provide a pedagogical and heuristic example in
Appendix \ref{app:sec:example}, in which the ES method works well, but
the cutoff method and the diagonal approximation fail manifestly.

\subsection{Bayesian method}
\label{subsec:bayes}

When we obtain the fitting function, we use the staggered chiral
perturbation theory to expand it in powers of $p^2$, $a^2$, and $m_q$.
In the series expansion, we must truncate the higher order terms
because we cannot include an arbitrary number of terms in the fitting
function.
One constraint is that we have only 4 data points of $B_K$ for the
SU(2) analysis.
Hence, the fitting function can have at most 3 unknown parameters, if
we want to perform the normal least $\chi^2$ fitting based on the
multivariate statistics theory.
This means that we can include all the next to leading order (NLO)
terms and one additional term at the next to next to leading order (NNLO).
It is the fitting function of Eq.~\eqref{eq:fit-func-1} that we obtain
following this premature logical path.
This looks fine as long as the truncated terms at the higher order are 
under control such that the full covariance fitting works well.
However, in our case of the SU(2) analysis on $B_K$, the full
covariance fitting fails manifestly because the data have such a high
precision that the truncated terms of the higher order are
required to fit the data.
We cannot add higher order terms to the fitting function
in a normal sense of the multivariate statistical theory.
Hence, the situation is checkmate as it is.
The question is how we can get out of this trap.
A natural answer is the Bayesian method \cite{Lepage-2001,sivia-2006}.

In the Bayesian method, the prior condition behaves in the fitting as
if it is one of the data points as explained in the previous
subsection.
Hence, it is possible to add $n$ higher order terms as long as we
impose $m$ prior conditions on the fitting with $n \le m$.
In practice, we impose the same number of prior conditions as that
of the higher order terms added to the fitting function as follows.
The fitting function has three additional terms at higher order:
\begin{eqnarray}
f_\text{th}^{\text{B}}(X)
  &=& f_\text{th}(X)  
      + c_4^b ~ X^2 \left(\ln X \right)^2
      + c_5^b ~ X^2 \left(\ln X \right)
      + c_6^b ~ X^3
\,.
\label{eq:f_th-bayes}
\end{eqnarray}
We impose the prior conditions on the fitting through the
prior probability as in the previous subsection:
\begin{equation}
\chi^2_\textrm{prior} =
 \sum_{k=4}^{6} \frac{({c_k^b} - a_{k;B})^2}{\sigma_{k;B}^2}
\end{equation}
Since we know that $c_k^b = 0 \pm 1$, we may choose $a_{k;B} =0$ and
$\sigma_{k;B} = 1$.
In the Bayesian method, we use $\chi^2_\textrm{aug}$ instead of $\chi^2$,
in the analysis, which is defined as 
\begin{eqnarray}
L &\equiv& \log(P(A|\mathbb{D},I)) \\
\chi^2_\textrm{aug} &=& (-2) L = \chi^2 + \chi^2_\textrm{prior}
\end{eqnarray}
where $P(A|\mathbb{D},I)$ is the posterior pdf \cite{sivia-2006}.
The Bayesian principle is that we determine the fitting
parameters such that they maximize the posterior pdf or
minimize $\chi^2_\textrm{aug}$.
The main advantage of the Bayesian method is that it allows for
probability interpretation and model selection as explained
in Ref.~\cite{sivia-2006}.
\begin{figure}[tbp]
\centering
  \includegraphics[width=20pc]
  {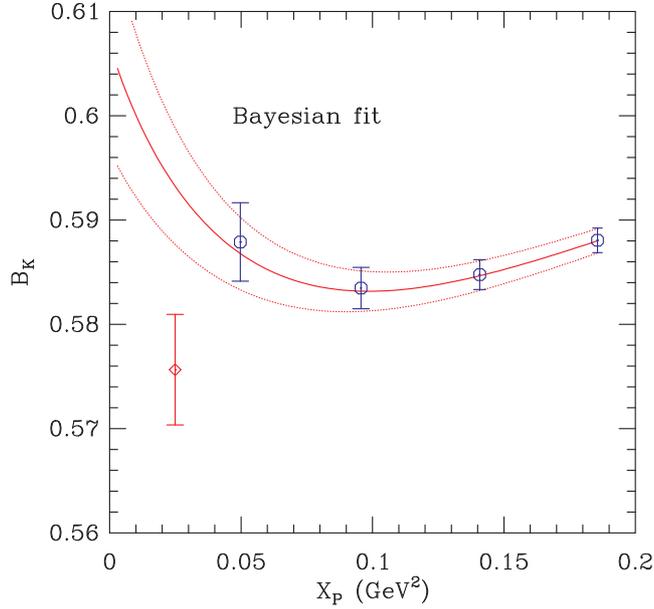}
  \caption{ $B_K(1/a)$ vs. $X_P$ on the C3 ensemble.  All the
    parameters are the same as in
    Figure \protect\ref{fig:diag-C3}. Here, the red line represents the
    results of the Bayesian method. 
  }
  \label{fig:bayes-C3}
\end{figure}
In Figure \ref{fig:bayes-C3}, we show the fitting results obtained using the
Bayesian method.
The fitting has $\chi^2_\textrm{aug} = 1.09(81)$.
The effective number of the data points is $\tilde{D} = 4 + 3 = 7$ and
the number of the unknown parameters is $\tilde{P} = 6$.
Hence, the effective number of degrees of freedom is 
$\tilde{d} = \tilde{D} - \tilde{P} = 1 = d$, the same as the full
covariance fitting.
\begin{table}[tbp]
\centering
\begin{tabular}{| c | c c c c |}
\hline
$i$ & 1 & 2 & 3 & 4 \\
\hline
$b_i$                 & 1.020(5)   & 0.5659(13) & 0.1060(3)    & 0.01442(3)    \\
$a_i$                 & 1.021(4)   & 0.5655(14) & 0.1061(3)    & 0.01442(3)    \\
$10^5 \cdot \delta_i$ & 110(41)    & 36.5(136)  & 5.00(187)    & 0.147(55)     \\
$\beta_i$             & 0.7583(16) & 0.2334(16) & 0.008194(69) & 0.0001515(12) \\
\hline
\end{tabular}
  \caption{ Eigenmode decomposition of $f_\text{th}^\text{B}$ for the
    fitting with the Bayesian method.}
  \label{tab:f-bayes}
\end{table}
In Table \ref{tab:f-bayes}, we decompose the fitting function
$f_\text{th}^\text{B}$ obtained using the Bayesian method in terms of
the eigenmodes of the full covariance matrix.
The fitting looks fine to our eyes.

A major advantage of the Bayesian method is that it does NOT touch the
covariance matrix at all unlike the cutoff method and the diagonal
approximation.

In the Bayesian method, we need to gauge the sensitivity of the
fitting to the prior condition.
In our case, we change the prior condition as follows:
\begin{equation}
\sigma_{k;B} = 1 \rightarrow 2
\end{equation}
while we keep $a_{k;B}$ unchanged.
The fitting results are changed as follows.
\begin{eqnarray}
B_K &=& 0.5757(53) \rightarrow 0.5772(58) 
\\
\chi^2_\textrm{aug} &=& 1.09(81) \rightarrow 0.31(23)
\end{eqnarray}
The mean value of $B_K$ is shifted by $0.28\sigma$ while
the error bar increases by 9\%.
Hence, the difference between the ES method and the Bayesian method is
$0.18\sigma$, which is smaller than the systematic error due to the
ambiguity in the prior condition.
Since both the ES and Bayesian methods are based on the Bayes theorem,
they are equivalent to each other in that sense.
When we obtain the higher order terms in Eq.~\eqref{eq:f_th-bayes}, we
use the continuum chiral perturbation theory but not the staggered
chiral perturbation theory.
Hence, the functional form of each higher order term is approximate
and not exact.
From this standpoint, we cannot claim that the Bayesian method is better
than the ES method.
Therefore, we decide to quote the difference between the results of ES and
Bayesian methods as our systematic error due to the ambiguity in the
covariance fitting in the SW-2, and to choose the results of the Bayesian 
method as the central value.

\section{Error analysis of the covariance matrix}
\label{sec:err-cov-mat}
It is true that elements of the covariance matrix may, in general,
have significantly larger errors than those of the small eigenvalues.
However, what makes the significant difference in fitting is the small
eigenvalues and the corresponding eigenmodes.
Hence, we focus on the error analysis of the small eigenvalues.
%

\begin{figure}[tbp]
\centering
  \includegraphics[width=20pc]{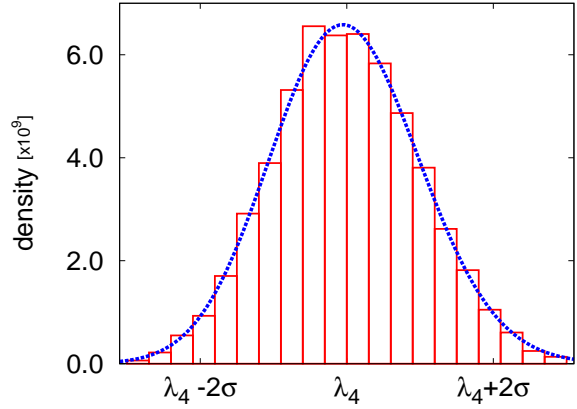}
  \caption{ Histogram of the $\lambda_4$ eigenvalue. The histogram has
    been obtained using the bootstrap method. The number of the
    bootstrap Monte Carlo samples is 10,000. The normalization of the
    histogram is adjusted such that the total probability is one. The
    blue curve represents the $\Gamma$ distribution. The details on 
    the $\Gamma$ distribution function are explained in Appendix
    \ref{app:sec:stat-anal}.  }
  \label{fig:hist-l4}
\end{figure}

The probability distribution of the small eigenvalues is the gamma
($\Gamma$) distribution, which is proved in Appendix
\ref{app:sec:stat-anal}.
Since the $\Gamma$ distribution is different from the normal
distribution, it is important to check whether our data respects the
$\Gamma$ distribution or not.
In Figure \ref{fig:hist-l4}, we show the probability density of the 
$\lambda_4$ eigenvalue in the format of histogram.
The blue curve in the plot corresponds to the $\Gamma$ distribution
function.
We find that the data respects the $\Gamma$ distribution very well.

In Table \ref{tab:err-anal-eigen}, we present the error of the
eigenvalues. The jackknife method is not very sensitive to the
left-right asymmetry of the distribution and provides a rough 
estimate of the errors. However, the bootstrap method is by nature
sensitive to the asymmetry. In the table, we show the jackknife
error as well as the left \& right errors of the bootstrap method.
As one can see, the left errors are consistently smaller than
the right errors, which indicates that the probability follows
the $\Gamma$ distribution.
\begin{table}[tbp]
\centering
\begin{tabular}{| c | c | c c c c |}
\hline
$i$ & scale & $\lambda_i$ & $\sigma_i(\text{jk})$ & 
$\sigma_i^L(\text{bs})$ & $\sigma_i^R(\text{bs})$ \\
\hline
1 & $10^{-6}$  & $19.51$ & 1.13(10) & 1.08(9)  & 1.16(11) \\
2 & $10^{-7}$  & $19.24$ & 1.08(7)  & 1.05(7)  & 1.11(8)  \\
3 & $10^{-9}$  & $75.79$ & 4.35(35) & 4.18(33) & 4.47(37) \\
4 & $10^{-11}$ & $110.9$ & 5.97(39) & 5.79(38) & 6.11(42) \\
\hline
\end{tabular}
  \caption{ Error analysis of eigenvalues.
    The scale represents the overall multiplication factor.
    $\sigma_i$ is the error of the $\lambda_i$. 
    The ``jk'' and ``bs'' index represent 
    the jackknife method and the bootstrap method, respectively.
    $\sigma_i^L$ and $\sigma_i^R$ represents the left error
    and right error.}
  \label{tab:err-anal-eigen}
\end{table}

\section{Conclusion \label{sec:conclude}}
Here, we address an issue of covariance fitting on the highly
correlated data: a general question frequently asked in the lattice
QCD community.
As an example, we have chosen the 4X3Y-NNLO fit of the $B_K$ data
based on the SU(2) staggered chiral perturbation theory explained in
SW-1.
It turns out that the smallest eigenvalue of the covariance matrix
leads to a extremely poor fitting.
If there exist very tiny eigenvalues in the full covariance matrix,
the small discrepancy between the fitting function and the data can be
dramatically amplified to make such a trouble that the fitting fails
in passing through the data points within the statistical uncertainty.
In order to get around the trouble, the lattice community have been
applying the diagonal approximation, the cutoff method, the modified
covariance method to the data analysis.
All of these poor prescriptions modify the covariance matrix in one
way or another and so lose the physical interpretation of $\chi^2$
completely.
Hence, we have been searching for a possible method which does not
touch the covariance matrix and allow for the physical interpretation
of the $\chi^2$.
A natural prescription which satisfies our requirements turns out to
be the Bayesian method and its variations.

In this paper, we suggest a new proposal: the eigenmode shift (ES)
method.
In this method, we shift the fitting function by a negligibly tiny
amount, which allows the full covariance fitting.
Note that we do not need to modify the covariance matrix at all in the
ES method, and, in addition, $\chi^2_\textrm{aug}$ has a physical
meaning based on the Bayesian probability interpretation.

Another good approach is the Bayesian method in which we add as many
higher order terms to the fitting function as the prior conditions
such that the fitting works well ({i.e.}  the
$\chi^2_\textrm{aug}$ has a reasonable value and the fitting looks
fine to our eyes).
One ambiguity in this method is our choice of higher order terms.
In our example of $B_K$, we use the continuum chiral perturbation
theory to obtain the functional form of each higher order term.
Since it is a kind of approximation and not exact, we cannot claim
that the Bayesian method is better than the ES method.

Our final suggestion is that it might be a good choice if one can try
both the ES and Bayesian methods in the data analysis and quote the
difference as the systematic uncertainty due to an ambiguity in the
covariance fitting.
We apply this approach to the error analysis in SW-2.

In order to help readers to digest the main points of this paper,
in Appendix \ref{app:sec:example}, we provide a pedagogical example of
data analysis, in which the diagonal approximation and the cut-off
method fail in fitting manifestly, but the ES method and the Bayesian
method work very well. This exemplifies an odd truth that
the conventional wisdom in the diagonal approximation and the cut-off
method might be falling apart in some cases.

\section*{Acknowledgements}
C.~Jung is supported by the US DOE under contract DE-AC02-98CH10886.
The research of W.~Lee is supported by the Creative Research
Initiatives program (No.~2012-0000241) of the NRF grant funded by the
Korean government (MEST).
Computations for this work were carried out in part on QCDOC computers
of the USQCD Collaboration at Brookhaven National Laboratory, and in
part on the DAVID GPU clusters at Seoul National University.
The USQCD Collaboration are funded by the Office of Science of the
U.S. Department of Energy.
W.~Lee acknowledges support from the KISTI supercomputing
center through the strategic support program (No.~KSC-2011-G2-06).

\appendix

\section{Equivalence of cutoff method and unconstrained ES method}
\label{app:sec:equiv_cutoff_ES}
Unconstrained ES method is the ES method whose shifting parameter, 
$\eta$, is not constrained by the Bayesian prior.
It is the same as the ES method whose prior condition is set to
$\sigma_\eta = \infty$.
In this section, we would like to prove that the unconstrained ES
method is equivalent to the cutoff method.

The $\chi^2$ of the cutoff method can be written in the following form:
\begin{equation}
\chi^2_\text{cut} = \langle\bar{y}-f | C'^{-1} | \bar{y}-f\rangle
\,.
\end{equation}
Here, $f$ is a vector representing the fitting function value,
\begin{equation*}
f_i = f_\text{th}(X_i)
\,,
\end{equation*}
$\bar{y}$ is the $D$-dimensional vector of average data points
and $C'^{-1}$ is the inverse covariance matrix of the cutoff method.
If $R$ number of eigenmodes, denoted by $S+1 \le i \le D$ with $S \equiv
D-R$, are removed from the covariance matrix by the cutoff method,
$C'^{-1}$ can be written as follows,
\begin{eqnarray}
\label{eq:cov_cutoff_eig}
C'^{-1} &=& C^{-1}
 - \sum_{i=S+1}^D \frac{1}{\lambda_{i}} |v_i\rangle \langle v_i|
  \nonumber \\
  &=& \sum_{i=1}^{S} \frac{1}{\lambda_{i}} |v_i\rangle \langle v_i|
\\
C^{-1} &=& \sum_{k=1}^D \frac{1}{\lambda_k} | v_k \rangle \langle v_k |
  \nonumber
\,.
\end{eqnarray}
Using the eigenmode decomposition given in 
Eq.~\eqref{eq:cov_cutoff_eig}, we can rewrite $\chi^2_\text{cut}$ as
\begin{equation}
\label{eq:chisq_cut}
\chi^2_\text{cut} = \sum_{i=1}^S \frac{1}{\lambda_i}
 \left[ \langle\bar{y}-f | v_i \rangle \right]^2
\,,
\end{equation}
while $\chi^2$ is defined by
\begin{eqnarray}
\chi^2 &=& \langle\bar{y}-f | C^{-1} | \bar{y}-f\rangle 
\nonumber \\
 &=& \sum_{i=1}^D \frac{1}{\lambda_i}
 \left[ \langle\bar{y}-f | v_i \rangle \right]^2
\,.  
\end{eqnarray}
Here, $C^{-1}$ is the inverse of the full covariance matrix.

The $\chi^2$ of the unconstrained ES method can be written in the 
following form:
\begin{equation}
\chi^2_\text{UES} = \langle\bar{y}-f' | C^{-1} | \bar{y}-f'\rangle
\,,
\end{equation}
where $f'$ is defined by
\begin{equation}
f' = f_\text{th}(X) + \sum_{i=S+1}^D \eta_{i} v_{i}
\,.
\end{equation}
The $\chi^2_\text{UES}$ can be expanded as follows,
\begin{eqnarray}
\chi^2_\text{UES} 
&=& \chi^2 - 2\sum_{i=S+1}^D \frac{1}{\lambda_{i}}
 \eta_{i}\langle\bar{y}-f | v_{i}\rangle
 + \sum_{i=S+1}^D \frac{1}{\lambda_{i}} \eta_{i}^2  \nonumber \\
&=& \chi^2 - \sum_{i=S+1}^D \frac{1}{\lambda_{i}}
 \left[ \langle\bar{y}-f | v_{i}\rangle \right]^2 
  + \sum_{i=S+1}^D \frac{1}{\lambda_{i}}
 \Big[ \eta_{i} - \langle\bar{y}-f | v_{i}\rangle 
 \Big]^2 \nonumber \\
&=& \chi_\text{cut}^2 + \sum_{i=S+1}^D \frac{1}{\lambda_{i}}
 \Big[ \eta_{i} - \langle\bar{y}-f | v_{i}\rangle 
 \Big]^2
 \label{eq:expand_ues}
\,.
\end{eqnarray}
Minimizing $\chi^2_\text{UES}$ yields
\begin{eqnarray}
\eta_{i} 
 &=& \langle\bar{y}-f | v_{i}\rangle 
\,,
\end{eqnarray}
and it removes the last term in Eq.~\eqref{eq:expand_ues}.
Hence, the minimizing $\chi^2_\text{cut}$ gives the same results
as those of the minimizing $\chi^2_\text{UES}$.
This proves our claim.

\section{Probability distribution of the minimized $\chi^2$}
\label{app:sec:dist_chisq}
\subsection{Distribution of $\chi^2$ for the full covariance fitting}
\label{app:subsec:dist_chisq_full}
As an introduction, we derive the probability distribution
of the minimized value of $\chi^2$ for the full covariance fitting.
Here, we assume that we have large number of data samples so that 
the sample covariance matrix is well determined.
Generally, the fitting function is inaccurate and the inaccuracy
leads us to the non-central $\chi^2$-distribution.
The non-central $\chi^2$-distribution is defined as follows.

Let us consider a set of independent random variables,
\begin{equation*}
\{X_1, X_2, X_3, \cdots, X_k \}
\,.
\end{equation*}
The random variables, $X_i$, follow the normal distribution with mean 
$\mu_i$ and variance $1$, denoted by $X_i \sim \mathcal{N}(\mu_i, 1)$.
If we define a new random variable $Q$ as follows,
\begin{equation}
Q = \sum_{i=1}^k X_i^2
\,,
\end{equation}
then $Q$ is distributed according to the non-central 
$\chi^2$-distribution with degrees of freedom $k$ and with 
non-centrality parameter
\begin{equation}
\kappa = \sum_{i=1}^k \mu_i^2
\,.
\end{equation}

The $\chi^2$ of the full covariance fitting can be written in the 
following form:
\begin{equation}
\label{eq:chisq_full_1}
\chi^2 = \langle\bar{y}-f | C^{-1} | \bar{y}-f\rangle
\,.
\end{equation}
Here, $f$ is a vector representing the fitting function value,
\begin{equation*}
f_i = f_\text{th}(X_i)
\,,
\end{equation*}
$\bar{y}$ is the $D$-dimensional vector of average data points
and $C^{-1}$ is the inverse of the covariance matrix.
The expectation value and covariance of the vector $\bar{y} - f$ are
\begin{eqnarray}
&&\mathcal{E}[\bar{y} - f] = \phi \label{eq:err_fit_ftn_phi}\\
&&\mathcal{C}\big[\bar{y} - f\big] = 
\mathcal{C}\big[\bar{y}\big] = C
\,.
\end{eqnarray}
Here, the vector $\phi$ represents the error of the fitting function.
If the fitting function is exact, $\phi = 0$.
According to the multi-dimensional central limit theorem, the average
of data, $\bar{y}$, is distributed as multivariate normal distribution.
Hence, $(\bar{y} - f)$ follows the multivariate normal distribution 
with mean vector $\phi$ and the covariance matrix $C$,
$(\bar{y} - f) \sim \mathcal{N}(\phi, C)$.

Let $M$ be a non-singular square matrix satisfying
\begin{equation}
MCM^T = I
\,,
\end{equation}
where $I$ is an identity matrix.
If we define a new vector $Y$ by
\begin{equation}
Y = M(\bar{y} - f)
\,,
\end{equation}
then the expectation value and covariance of $Y$ is
\begin{eqnarray}
\mathcal{E}[Y] &=& M\phi \\
\mathcal{C}\big[Y\big] &=& MCM^T = I
\,.
\end{eqnarray}
Since $M$ is non-singular, the transformed vector $Y$ is also 
distributed according to the multivariate normal distribution
(Proof is given in Ref.~\cite{anderson-2003}).
Hence, $Y$ is distributed according to the multivariate normal
distribution with mean $M\phi$ and covariance matrix $I$.
Note that identity covariance matrix implies mutual independence 
of $Y_i$.

In terms of $Y$, the $\chi^2$, given in Eq.~\eqref{eq:chisq_full_1}, 
can be rewritten by
\begin{eqnarray}
\chi^2 &=& \langle Y | Y \rangle \nonumber \\
 &=& \sum_{i=1}^D Y_i^2
\,.
\end{eqnarray}
Because $Y_i$ are independently distributed as normal distribution
with mean $[M\phi]_i$ and variance $1$, $\chi^2$ is distributed
according to the non-central $\chi^2$-distribution.
If the number of fitting parameters is $P$, degrees of freedom of 
the distribution is $D-P$.
The non-centrality parameter of the distribution is
\begin{eqnarray}
\kappa &=& \sum_{k=1}^D \Big( [M\phi]_k \Big)^2 \nonumber \\
 &=& (M\phi)^T (M\phi) = \phi^T M^T M \phi = \phi^T C^{-1} \phi
\,,
\end{eqnarray}
where we used $M^TM = C^{-1}$ at the last equality.

\subsection{Distribution of $\chi^2$ for the cutoff method}
\label{app:subsec:dist_chisq_cutoff}
As described in Appendix~\ref{app:sec:equiv_cutoff_ES}, 
the $\chi^2$ of the cutoff method can be written in the following form:
\begin{equation}
\chi^2_\text{cut} = \langle\bar{y}-f | C'^{-1} | \bar{y}-f\rangle
\,,
\end{equation}
where $C'^{-1}$ is the inverse covariance matrix of the cutoff 
method, which can be written by
\begin{eqnarray}
C'^{-1} &=& C^{-1}
 - \sum_{i=S+1}^D \frac{1}{\lambda_{i}} |v_{i}\rangle \langle v_{i}|
  \\
C^{-1} &=& \sum_{k=1}^D \frac{1}{\lambda_k} | v_k \rangle \langle v_k |
  \nonumber
\,.
\end{eqnarray}
Here, $R$ is the number of removed eigenmodes, and $S=D-R$.
Let $M$ be a non-singular square matrix satisfying
\begin{equation}
MCM^T = I
\,,
\end{equation}
where $I$ is an identity matrix.
Using the eigenmode decomposition, it is 
\begin{eqnarray}
M &=& \sum_{k=1}^D \frac{1}{\sqrt{\lambda_k}} |e_k\rangle\langle v_k| \\
M^{-1} &=& \sum_{k=1}^D \sqrt{\lambda_k} |v_k\rangle\langle e_k|
\,,
\end{eqnarray}
where $e_k$ is a unit vector in $k$-direction.
In terms of the new vector $Y = M(\bar{y} - f)$, the $\chi^2_\text{cut}$ 
becomes
\begin{eqnarray}
\chi^2_\text{cut} 
 &=& \langle Y | (M^T)^{-1} C'^{-1} M^{-1} | Y\rangle \nonumber \\
 \label{eq:chisq_cut_1}
 &=& \sum_{k=1}^D Y_k^2 - \sum_{i=S+1}^D Y_{i}^2 \\
 &=& \sum_{k=1}^S Y_k^2
\,.
\end{eqnarray}
Here, we used the following relation,
\begin{equation}
(M^T)^{-1} C'^{-1} M^{-1} 
 = I - \sum_{i=S+1}^D | e_{i} \rangle \langle e_{i} |
\,.
\end{equation}
As mentioned in Appendix~\ref{app:subsec:dist_chisq_full}, $Y_i$ are
independently distributed as normal distribution with mean $[M\phi]_i$
and variance 1, $Y_i \sim \mathcal{N}\big([M\phi]_i, 1\big)$.
Hence, Eq.~\eqref{eq:chisq_cut_1} shows that $\chi^2_\text{cut}$ is
distributed according to the non-central $\chi^2$-distribution with
degrees of freedom $D-P-R$ and with non-centrality parameter
\begin{eqnarray}
\kappa 
 &=& \sum_{k=1}^D \Big( [M\phi]_k \Big)^2 
  - \sum_{i=S+1}^D \Big( [M\phi]_{i} \Big)^2 \nonumber \\
 &=& \sum_{k=1}^S \frac{1}{\lambda_k}
 \Big[ \langle v_k | \phi \rangle \Big]^2
 \label{eq:kappa_cutoff}
\,,
\end{eqnarray}
where $\phi$, which is a vector representing the error of fitting 
function, is defined in Eq.~\eqref{eq:err_fit_ftn_phi}.
Here, $D$ is the number of data points, $P$ is the number of fitting
parameters and $R$ is the number of removed eigenmodes.
If we assume that the fitting function is exact $(\phi = 0)$, 
$\chi^2_\text{cut}$ follows the $\chi^2$-distribution with degrees
of freedom: $S-P = D-P-R$.
%


%
\subsection{Distribution of $\chi^2$ for the ES method}
\label{app:subsec:dist_chisq_ES}

In Appendix~\ref{app:sec:equiv_cutoff_ES}, we show that unconstrained
ES method is the same as the cutoff method.
The probability distribution of the minimized value of $\chi^2$ in
cutoff method is derived in Appendix~\ref{app:subsec:dist_chisq_cutoff}
and it is the $\chi^2$-distribution with degrees of freedom $D-P-R$,
if we assume that the fitting function is good enough.
Hence, the probability distribution of the minimized value of $\chi^2$ 
in unconstrained ES method is the $\chi^2$-distribution with degrees of
freedom $D-P-R$.
Here, $R$ is the number of shifting eigenvectors that modify the 
fitting function.

The deformation of the degrees of freedom in unconstrained ES method 
can be considered as a consequence of adding new fitting parameters
that control the shifting eigenvectors.
We add $R$ number of shifting parameters, $\{\eta_i\}$ with
$i=1,2,\ldots,R$, and it increases the number of fitting parameters to
$P+R$.
As a result, the degrees of freedom becomes $D-P-R$.

In the normal ES method, we constrain the shifting parameters by
Bayesian method.
Hence, distribution of the augmented $\chi^2$ in ES method can be 
interpreted that of in the Bayesian constrained fitting.

By the Bayes theorem,
\begin{equation}
P(A|\mathbb{D},I) \propto P(\mathbb{D} | A, I) \times P(A | I)
\end{equation}
Here, $P(\mathbb{D} | A, I)$ is the likelihood function, which
can be expressed as
\begin{equation}
P(\mathbb{D} | A, I) \propto \exp\left(-\frac{\chi^2}{2}\right)
\end{equation}
In addition, let us assume that the prior probability $P(A | I)$ can
be written as
\begin{equation}
P(A | I) \propto \exp\left(-\frac{\chi^2_\textrm{prior}}{2}\right)
\end{equation}
The effective number of degrees of freedom in the ES method is
$\tilde{d} = \tilde{D} - \tilde{P} = D - P= d$, where $\tilde{D} = D +
R$ and $\tilde{P} = P + R$ since there are $R$ number of the prior
conditions imposed on $\eta_i$ and there exists as much increase in
the number of fitting parameters by $R$.
Hence, the $\chi^2_\textrm{aug} = \chi^2 + \chi^2_\textrm{prior}$
must follow the normal $\chi^2$ distribution with 
$\chi^2_\textrm{aug} = \tilde{d} \pm \sqrt{2 \tilde{d}} = d \pm \sqrt{2 d}$.
Therefore, the $\chi^2_\textrm{aug}$ of the ES method follows the same 
normal $\chi^2$ distribution as that of the full covariance fitting.

\section{An example of fitting with random data}
\label{app:sec:example}
In this section, we will show an example to see how the fitting methods,
given in section~\ref{sec:pres}, work.

First, let us explain how the data has been generated.
The true mean $\mu_i$ of the data is 
\begin{equation}
\mu_i = f_\textrm{true}(x_i) = \frac{1}{1-x_i}
\end{equation} 
We choose 7 data points ($D=7$) such that 
$x_i = 0.2 \times \dfrac{(i-1)}{6}$ with $i=1,2,3,\ldots,7$.
We also choose the eigenvalues of the true covariance matrix as
follows,
\begin{equation}
\lambda^\Gamma_i = 10^{-\frac{2}{3}(i-1)}
\end{equation}
for $i = 1,2,3,\ldots,7$.
Then, we generate the random numbers to construct the eigenvectors
which are orthonormal to each other by construction.
Hence, we can obtain the true covariance matrix $\Gamma_{ij}$
from the eigenvalues and eigenvectors.
Note that the covariance matrix for the sample mean $C_{ij}$ is
related to the $\Gamma_{ij}$ as follows:
\[
C_{ij} = \frac{1}{N} \Gamma_{ij}
\]
Hence, the eigenvalues of $C_{ij}$ is smaller by a factor of $N$
as follows.
\[
\lambda^C_i = \frac{1}{N} \lambda^\Gamma_i
\]
For the notational convenience, we drop out the superscipt for the
eigenvalues in the discussion in this section.

Next, we generate the data $y_i$ following the probability
distribution: $Y \sim \mathcal{N}(\mu,\Gamma)$.
Here, we use the numerical algorithm introduced in Ref.~\cite{NR-7-4}
to generate the data of $y_i$ according to the pdf:
\begin{eqnarray}
& & P(y|\mu, \Gamma) = Z
\exp[ -\frac{1}{2} (y-\mu) \cdot \Gamma^{-1} \cdot (y-\mu) ]
\\ & & 
Z = \frac{1}{(2\pi)^D \det(\Gamma)^{1/2}}
\end{eqnarray}
In this way, we collect 1000 random data samples for $y_i$
({i.e.} $N=1000$).

We know that the true fitting function can be rewritten as follows,
\begin{equation}
f_\textrm{true}(x) = \frac{1}{1-x} 
= 1 + x + x^2 + \cdots + x^n + \cdots
\end{equation}
We choose the trial fitting function as follows:
\begin{equation}
f_\textrm{trial}(x) = c_1 + c_2 x + c_3 x^2
\end{equation}
In addition, we also choose the trial fitting function for the
Bayesian method as
\begin{equation}
f_\textrm{Bayes}(x) = c_1 + c_2 x + c_3 x^2 + c^{b}_4 x^3 \,,
\end{equation}
where we will impose the prior condition on $c^{b}_4$.

We fit the data using the fitting methods such as the full covariance 
fitting, diagonal approximation, cut-off method, modified covariance
method, ES method, and Bayesian method. 
In the cut-off method, we remove lowest two eigenmodes by imposing
the cut-off ($\lambda_i/\lambda_1 \ge 1.0 \times 10^{-3}$),
where $\lambda_1$ is the largest eigenvalue.
In the ES method, we introduce two shift parameters $\eta_1$ and
$\eta_2$ in the direction of the two smallest eigenmodes.
We impose the following prior condition on $\eta_i$.
\begin{eqnarray}
\eta_i &\sim& \mathcal{N}(a_i, \sigma_i^2) \\
a_i &=& 0 \nonumber \\
\sigma_i &=& 0.001 \nonumber
\end{eqnarray}
Here, the highest terms of the truncated are of order $\mathcal{O}(x^3)$.
By assuming that the coefficient is $\mathcal{O}(1)$, we can 
estimate the size of truncated terms approximately as
\[
\sigma_i \approx 1.0 \times x^3 \approx 1.0 \times (0.1)^3 = 0.001 \,,
\]
where we take the average of $x_i$ as the $x$ value.

In the Bayesian method, we impose the following prior condition
on $c_4^b$.
\begin{eqnarray}
c_4^b &\sim& \mathcal{N}(a_4, \sigma_4^2) \\
a_4 &=& 0 \nonumber \\
\sigma_4 &=& 1 \nonumber
\end{eqnarray}
%

%
%
%
\begin{figure}[tbp]
\centering
  \includegraphics[width=25pc]{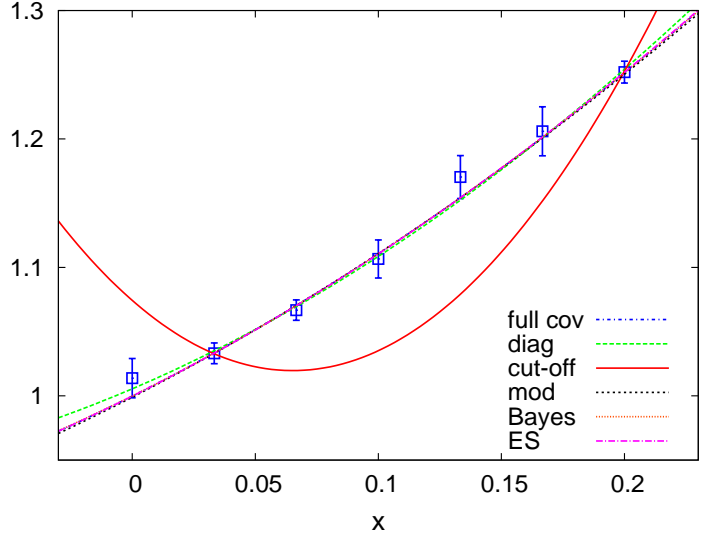}
  \caption{Comparison of various fitting methods: 
    full covariance fitting (full cov), diagonal approximation (diag), 
    cut-off method (cut-off), modified covariance matrix method (mod), 
    Bayesian method (Bayes) and eigenmode shift method (ES).}
  \label{fig:cmp-fit}
\end{figure}
In Figure \ref{fig:cmp-fit}, we show the fitting results obtained
using various fitting methods.
As one can see in the plot, it is very clear that there is something
seriously wrong with the cut-off method.
In Table \ref{tab:sample-fit}, we present results for various fitting
methods.
The results of the full covariance fitting is what we want to achieve
in the fitting business ultimately.
As one can see in the table, the diagonal approximation and the
cut-off method fail, manifestly. 
In other words, errors of the $c_i$ parameters are dramatically larger
than those of the full covariance fitting, which indicates a symptom
of a complete failure in fitting.
In addition, we notice that both the ES method and the Bayesian method
work very well in good agreement with the expectation, while the ES
method looks marginally better in this example.
In the case of the ES method, we obtain the following results for the
shift parameters:
\begin{eqnarray*}
\eta_1 &=& 0.00015 \pm 0.00154 \\
\eta_2 &=& -0.00023 \pm 0.00039\,,
\end{eqnarray*}
which are highly self-consistent with the prior condition.
\begin{table}[tbp]
\centering
\begin{tabular}{| l | l l l | l |}
\hline
fit type & $c_1$       & $c_2$      & $c_3$     & $\chi^2/\text{dof}$  \\
\hline
full cov & 0.9998(19)  & 0.957(41)  & 1.51(24)  & 0.65(81) \\
diag     & 1.0055(105) & 0.818(211) & 2.13(97)  & 0.45(96) \\
cut-off  & 1.0743(808) & -1.67(285) & 12.8(122) & 0.70(118) \\
mod      & 0.9989(20)  & 0.981(47)  & 1.38(27)  & 1.9(26) \\
Bayes    & 0.9999(19)  & 0.950(42)  & 1.59(27)  & 0.63(79) \\
ES       & 0.9998(19)  & 0.954(42)  & 1.53(24)  & 0.60(76) \\
\hline
\end{tabular}
  \caption{Fitting results of various fitting methods:
  The fitting details are explained in the text.}
  \label{tab:sample-fit}
\end{table}

In this example, we learn the following lessons:
\begin{itemize}
\item In this example, the full covariance fitting works very
  well. Hence, we know the fitting results a priori.  In addition, we
  know the true answer to the fitting.
\item In this example, the diagonal approximation overestimates the
  error of the $c_1$ parameter by factor of 5. To make matters worse,
  the cut-off method overestimates the same quantity by factor of 40.
  This indicates that we should be very careful when we use these
  methods.  They are very likely to overestimate the errors, which
  could lead to a wrong answer as in this example. Hence, we do not
  recommend both the diagonal approximation and the cut-off method
  for the data analysis to aim at the high precision.
\item In this example, the ES method and the Bayesian method work very
  well in good agreement with the full covariance fitting, while the
  ES method is marginally better. Hence, we decide using these two
  methods to do the data analysis for $B_K$ as quoted in SW-2.
\end{itemize}

\section{Statistical analysis of the eigenvalues}
\label{app:sec:stat-anal}
\subsection{Distribution of the eigenvalues}
\label{app:subsec:dist-eig}
In this section, we will talk about the distribution of the eigenvalues 
of our covariance matrix. 
Let us consider a set of independent and identically distributed  
random variables : $\{X_1, X_2, \cdots, X_N\}$.
A sample variance can be defined by
\begin{equation}
s^2 = \frac{1}{N-1} \sum_{i=1}^{N} (X_i - \overline{X})^2,
\end{equation}
where $\overline{X}$ is an average of $X_i$ over $i$.
If $X_i$ are distributed according to the normal-distribution
with mean $\mu$ and variance $\sigma^2$, then $(N-1)s^2/\sigma^2$
is distributed according to the $\chi^2$-distribution with
degrees of freedom $N-1$ :
\begin{equation}
(N-1)\frac{s^2}{\sigma^2} 
  = \sum_{i=1}^{N} \frac{(X_i - \overline{X})^2}{\sigma^2}
  \sim \chi^2_{N-1}.
\end{equation}
Since we have $N=671$ gauge configurations averaged from 9 measurements
(see Table \ref{tab:para-C3}), we can assume that the data is 
distributed according to the normal-distribution.
Hence, the sample variance multiplied by a constant of our data is 
distributed according to the $\chi^2$-distribution with degrees of 
freedom $N-1$. 

If a random variable $Y$ is distributed according to the 
$\chi^2$-distribution with degrees of freedom $\nu$, then a new random 
variable multiplied by a positive constant $c$ is distributed according 
to the gamma-distribution ($\Gamma$-distribution) with shape parameter 
$k=\nu/2$ and scale parameter $\theta = 2c$:
\begin{eqnarray}
Y  &\sim& \chi^2_\nu, \nonumber \\
cY &\sim& \Gamma(k = \frac{\nu}{2}, \theta = 2c).
\end{eqnarray}
Hence, sample variance of normally distributed random samples
is distributed according to the $\Gamma$-distribution.

There is an orthogonal transformation of the basis vectors that makes
the covariance matrix diagonal with its diagonal elements equal to
the eigenvalues.
Note that a linear combination of normal random variables is also
normally distributed as proved in Ref.~\cite{anderson-2003}.  We may
regard the eigenvalues as a variance of normally distributed random
variables (data). 
Therefore, the eigenvalues of the covariance matrix are distributed
according to the $\Gamma(\frac{N-1}{2}, \theta_i)$ distribution.
If we assume that the expectation values of eigenvalues are
the measured eigenvalues, $\lambda_i$, it gives
\begin{equation}
  \theta_i = \frac{2}{N-1} \lambda_i,
\end{equation}
where we used a property of $\Gamma$-distribution that a mean value of
$\Gamma(k, \theta)$ is $k\theta$ \cite{schervish-1995,hogg-2005}.

The probability distribution function (pdf) of $\Gamma$-distribution
with shape parameter $k$ and scale parameter $\theta$ (denoted
$\Gamma(k, \theta)$) is
\begin{equation}
p(x;k,\theta) 
  = x^{k-1}\frac{e^{-x/\theta}}{\theta^k \Gamma(k)} 
  \qquad \text{ for } x \ge 0 \,.
\end{equation}
Here, $k$ and $\theta$ should be positive and $\Gamma(k)$ is a 
gamma function.
The mean and variance are $k\theta$ and $k\theta^2$, respectively.

\subsection{Error of eigenvalues}
In table \ref{tab:err-anal-eigen}, we presented the error of
eigenvalues measured by resampling method : the jackknife method and
the bootstrap method.
In this section, we will describe other possible methods that estimate
errors of eigenvalues.

The first method is to use properties of the $\Gamma$-distribution of
eigenvalues.
Using results of Appendix \ref{app:subsec:dist-eig}, we know that
the distribution of eigenvalues are $\Gamma$-distribution with shape
parameter $k=\frac{N-1}{2}$ and scale parameter $\theta_i =
\frac{2}{N-1}\lambda_i$.
Since the variance of the distribution is $k\theta_i^2$, the statistical
error of the eigenvalues are $\sqrt{k\theta_i^2}$.

The second method is using the error propagation formula. 
The error of eigenvalues basically originates from the error of 
the covariance matrix.
Hence, we can measure the fluctuation of the eigenvalues fluctuating
covariance matrix.
In other words, we obtain eigenvalues of new covariance matrices
fluctuated by
\begin{equation}
\Sigma_{ij} = C_{ij} + \delta_{ij},
\label{eq:new-cov-mat}
\end{equation}
where $\delta_{ij}$ is a Gaussian random noise.
Note that the elements of the covariance matrix are correlated, which 
means that the random noise should be generated containing the information
of the correlation. 
Hence, we generated the random noise following multivariate normal 
distribution with zero mean and given covariance, 
$\text{Cov}(C_{ij}, C_{kl})$.

Let us calculate the covariance matrix of the covariance matrix,
$\text{Cov}(C_{ij}, C_{kl})$.  
The covariance matrix defined in Eq.~\eqref{eq:cov_mat} can be 
rearranged as follows:
\begin{eqnarray}
C_{ij} &=& \frac{1}{N} \sum_{n} x_{ij}(n) = \bar{x}_{ij},
\\
x_{ij}(n) &=& \frac{1}{N-1} [y_i(n) - \bar{y}_i] [y_j(n) - \bar{y}_j]
\label{eq:x_ij}.
\end{eqnarray}
Hence, the covariance matrix of the covariance matrix can be defined
as follows,
\begin{eqnarray}
\text{Cov}(C_{ij},C_{kl}) 
 &=& \frac{1}{N(N-1)} \sum_n [x_{ij}(n) - \bar{x}_{ij}] 
  [x_{kl}(n) - \bar{x}_{kl}]
  \nonumber \\
 &=& \frac{1}{N(N-1)} \sum_n [x_{ij}(n) - C_{ij}] [x_{kl}(n) - C_{kl}].
\label{eq:cov-cov}
\end{eqnarray}
%

%
\begin{table}[tbp]
\centering
\begin{tabular}{| c | c | c c c c c |}
\hline
$i$ & scale & $\lambda_i$ & $\sigma_i(\text{JK})$ & 
 $\sigma_i(\text{GD})$ & $\sigma_i(\text{EP})$ & $\sigma_i(\text{FM})$ \\
\hline
1 & $10^{-6}$  & $19.51$ & 1.13 & 1.07 & 1.12 & 1.07 \\
2 & $10^{-7}$  & $19.24$ & 1.08 & 1.05 & 1.07 & 1.05 \\
3 & $10^{-9}$  & $75.79$ & 4.35 & 4.14 & 4.33 & 4.15 \\
4 & $10^{-11}$ & $110.9$ & 5.97 & 6.06 & 5.92 & 6.06 \\
\hline
\end{tabular}
  \caption{ Error of eigenvalues estimated by different methods.
    The scale represents the overall multiplication factor.
    $\sigma_i$ is the error of the $\lambda_i$. 
    The ``JK'', ``GD'', ``EP'' and ``FM'' index represent 
    the jackknife method, gamma-distribution method, 
    error propagation method and error propagation with fourth 
    moment method, respectively.
    Each $\sigma_i$ element has about 8\% error. 
    }
  \label{tab:err-eig}
\end{table}
%
%

The third method is using the fourth moment of the normal distribution.
It is well known that the 4th moment of normally distributed
random variables, $Z_i$, is
\begin{eqnarray}
\mathcal{E}\big[(Z_i - \mu_i)(Z_j - \mu_j)(Z_k - \mu_k)(Z_l - \mu_l)\big]
= \sigma_{ij}\sigma_{kl} + \sigma_{ik}\sigma_{jl} + \sigma_{il}\sigma_{jk},
\label{eq:4th-moment}
\end{eqnarray}
where $\mu_i$ are the true expectation values of $Z_i$ and
$\sigma_{ij}$ are the true covariance between $Z_i$ and $Z_j$ (See
Ref.~\cite{anderson-2003}).
Let us assume that $C_{ij}$ can be counted as a true covariance of mean
of two random variables in the limit of $N \rightarrow \infty$.
Then, $C_{ij}$ is related to $\sigma_{ij}$ in the limit of 
$N \rightarrow \infty$:
\[
C_{ij} = \frac{1}{N} \sigma_{ij}
\]
If we assume that $C_{ij}$ can be counted as a true covariance of mean
of two random variables, $\text{Cov}(C_{ij}, C_{kl})$ can be estimated
by
\begin{eqnarray}
\text{Cov}(C_{ij}, C_{kl})
 &=& \frac{1}{N} \mathcal{E}
  \bigg[ \left(\frac{(Z_i - \mu_i)(Z_j - \mu_j)}{N} - \mathcal{E}\big[C_{ij}\big]\right)
  \nonumber \\ & & \hspace{10mm}
  \times \left( \frac{(Z_k - \mu_k)(Z_l - \mu_l)}{N} - \mathcal{E}\big[C_{kl}\big]\right)
  \bigg]
\nonumber \\
 &=& \frac{1}{N}(C_{ik}C_{jl} + C_{il}C_{jk}),
\end{eqnarray}
where we used Eq.~\eqref{eq:4th-moment} to obtain the last equality.

The results of error analysis on the eigenvalues are summarized in
Table \ref{tab:err-eig}.
All the methods show the same results within statistical uncertainty.

\bibliographystyle{model1-num-names} 
\bibliography{ref} 

\begin{thebibliography}{27}
\expandafter\ifx\csname natexlab\endcsname\relax\def\natexlab#1{#1}\fi
\providecommand{\bibinfo}[2]{#2}
\ifx\xfnm\relax \def\xfnm[#1]{\unskip,\space#1}\fi
\bibitem[{Bae et~al.(2010)Bae, Jang, Jung, Kim, Kim, Kim, Kim, Lee, Sharpe, and
  Yoon}]{wlee-10-3}
\bibinfo{author}{T.~Bae}, \bibinfo{author}{Y.-C. Jang},
  \bibinfo{author}{C.~Jung}, \bibinfo{author}{H.-J. Kim},
  \bibinfo{author}{J.~Kim}, \bibinfo{author}{J.~Kim}, \bibinfo{author}{K.~Kim},
  \bibinfo{author}{W.~Lee}, \bibinfo{author}{S.~R. Sharpe},
  \bibinfo{author}{B.~Yoon},
\newblock \bibinfo{title}{{$B_K$ using HYP-smeared staggered fermions in Nf=2+1
  unquenched QCD}},
\newblock \bibinfo{journal}{Phys. Rev.} \bibinfo{volume}{D82}
  (\bibinfo{year}{2010}) \bibinfo{pages}{114509}.
\bibitem[{Bae et~al.(2011)Bae, Jang, Jung, Kim, Kim et~al.}]{Bae:2011ff}
\bibinfo{author}{T.~Bae}, \bibinfo{author}{Y.-C. Jang},
  \bibinfo{author}{C.~Jung}, \bibinfo{author}{H.-J. Kim},
  \bibinfo{author}{J.~Kim}, et~al.,
\newblock \bibinfo{title}{{Kaon $B$-parameter from improved staggered fermions
  in $N_f=2+1$ QCD}}  (\bibinfo{year}{2011}).
\bibitem[{Thacker and Lepage(1991)}]{Thacker:1990bm}
\bibinfo{author}{B.~Thacker}, \bibinfo{author}{G.~Lepage},
\newblock \bibinfo{title}{{Heavy quark bound states in lattice QCD}},
\newblock \bibinfo{journal}{Phys.Rev.} \bibinfo{volume}{D43}
  (\bibinfo{year}{1991}) \bibinfo{pages}{196--208}.
\bibitem[{Drummond and Horgan(1993)}]{Drummond:1992pg}
\bibinfo{author}{I.~Drummond}, \bibinfo{author}{R.~Horgan},
\newblock \bibinfo{title}{{Improved Langevin methods for spin systems}},
\newblock \bibinfo{journal}{Phys.Lett.} \bibinfo{volume}{B302}
  (\bibinfo{year}{1993}) \bibinfo{pages}{271--278}.
\bibitem[{Kilcup(1994)}]{kilcup-1994-1}
\bibinfo{author}{G.~Kilcup},
\newblock \bibinfo{title}{{Quenched staggered spectrum at beta = 6.0, 6.2 and
  6.4}},
\newblock \bibinfo{journal}{Nucl. Phys. Proc. Suppl.} \bibinfo{volume}{34}
  (\bibinfo{year}{1994}) \bibinfo{pages}{350--354}.
\bibitem[{Michael(1994)}]{michael-1994-1}
\bibinfo{author}{C.~Michael},
\newblock \bibinfo{title}{{Fitting correlated data}},
\newblock \bibinfo{journal}{Phys. Rev. D} \bibinfo{volume}{49}
  (\bibinfo{year}{1994}) \bibinfo{pages}{2616}.
\bibitem[{Michael and McKerrell(1995)}]{Michael:1994sz}
\bibinfo{author}{C.~Michael}, \bibinfo{author}{A.~McKerrell},
\newblock \bibinfo{title}{{Fitting correlated hadron mass spectrum data}},
\newblock \bibinfo{journal}{Phys.Rev.} \bibinfo{volume}{D51}
  (\bibinfo{year}{1995}) \bibinfo{pages}{3745--3750}.
\bibitem[{{Steven Gottlieb and W. Liu and R.L. Renken and R.L. Sugar and Doug
  Toussaint}(1988)}]{milc-1988-1}
\bibinfo{author}{{Steven Gottlieb and W. Liu and R.L. Renken and R.L. Sugar and
  Doug Toussaint}},
\newblock \bibinfo{title}{{Hadron masses with two quark flavors}},
\newblock \bibinfo{journal}{Phys. Rev. D} \bibinfo{volume}{38}
  (\bibinfo{year}{1988}) \bibinfo{pages}{2245}.
\bibitem[{Toussaint(1990)}]{toussaint-1990-1}
\bibinfo{author}{D.~Toussaint}, \bibinfo{title}{From Action to Answers},
  \bibinfo{publisher}{World Scientific}, \bibinfo{address}{Singapore},
  \bibinfo{year}{1990}. \bibinfo{note}{Page 121}.
\bibitem[{Anderson(2003)}]{anderson-2003}
\bibinfo{author}{T.~W. Anderson}, \bibinfo{title}{{An Introduction to
  Multivariate Statistical Analysis}}, Wiley Series in Probability and
  Statistics, \bibinfo{publisher}{Wiley Interscience}, \bibinfo{edition}{third}
  edition, \bibinfo{year}{2003}.
\bibitem[{Johnson and Wichern(2007)}]{johnson-2007}
\bibinfo{author}{R.~A. Johnson}, \bibinfo{author}{D.~W. Wichern},
  \bibinfo{title}{{Applied Multivariate Statistical Analysis}},
  \bibinfo{publisher}{Pearson Prentice Hall}, \bibinfo{edition}{sixth} edition,
  \bibinfo{year}{2007}.
\bibitem[{Lee and Sharpe(1999)}]{wlee-99}
\bibinfo{author}{W.~Lee}, \bibinfo{author}{S.~R. Sharpe},
\newblock \bibinfo{title}{{Partial Flavor Symmetry Restoration for Chiral
  Staggered Fermions}},
\newblock \bibinfo{journal}{Phys. Rev.} \bibinfo{volume}{D60}
  (\bibinfo{year}{1999}) \bibinfo{pages}{114503}.
\bibitem[{Aubin and Bernard(2003)}]{bernard-03}
\bibinfo{author}{C.~Aubin}, \bibinfo{author}{C.~Bernard},
\newblock \bibinfo{title}{{Pion and kaon masses in staggered chiral
  perturbation theory}},
\newblock \bibinfo{journal}{Phys. Rev. D} \bibinfo{volume}{68}
  (\bibinfo{year}{2003}) \bibinfo{pages}{034014}.
\bibitem[{Bazavov et~al.(2010)Bazavov, Toussaint, Bernard, Laiho, DeTar
  et~al.}]{milc-rmp-09}
\bibinfo{author}{A.~Bazavov}, \bibinfo{author}{D.~Toussaint},
  \bibinfo{author}{C.~Bernard}, \bibinfo{author}{J.~Laiho},
  \bibinfo{author}{C.~DeTar}, et~al.,
\newblock \bibinfo{title}{{Full nonperturbative QCD simulations with 2+1
  flavors of improved staggered quarks}},
\newblock \bibinfo{journal}{Rev. Mod. Phys.} \bibinfo{volume}{82}
  (\bibinfo{year}{2010}) \bibinfo{pages}{1349--1417}.
\bibitem[{de~Water and Sharpe(2006)}]{steve-06}
\bibinfo{author}{R.~V. de~Water}, \bibinfo{author}{S.~Sharpe},
\newblock \bibinfo{title}{{$B_K$ in staggered chiral perturbation theory}},
\newblock \bibinfo{journal}{Phys. Rev. D} \bibinfo{volume}{73}
  (\bibinfo{year}{2006}) \bibinfo{pages}{014003}.
\bibitem[{Schervish(1995)}]{schervish-1995}
\bibinfo{author}{M.~J. Schervish}, \bibinfo{title}{{Theory of Statistics}},
  \bibinfo{publisher}{Springer-Verlag}, \bibinfo{year}{1995}.
\bibitem[{Toussaint and Freeman(2008)}]{toussaint-08}
\bibinfo{author}{D.~Toussaint}, \bibinfo{author}{W.~Freeman},
\newblock \bibinfo{title}{{Sample size effects in multivariate fitting of
  correlated data}}  (\bibinfo{year}{2008}).
\bibitem[{Press et~al.(2007)Press, Teukosky, Vetterling, and
  Flannary}]{NR-2007-1}
\bibinfo{author}{W.~Press}, \bibinfo{author}{S.~Teukosky},
  \bibinfo{author}{W.~Vetterling}, \bibinfo{author}{B.~Flannary},
  \bibinfo{title}{Numerical Recipes}, \bibinfo{publisher}{Cambridge University
  Press}, \bibinfo{address}{New York}, \bibinfo{year}{2007}.
  \bibinfo{note}{Chapter 15, section 4}.
\bibitem[{Bailey et~al.(2010{\natexlab{a}})}]{ref:fnal-2010-1}
\bibinfo{author}{J.~Bailey}, et~al.,
\newblock \bibinfo{title}{{Semileptonic decays of K and D mesons in 2+1 flavor
  QCD}},
\newblock \bibinfo{journal}{PoS} \bibinfo{volume}{LATTICE 2010}
  (\bibinfo{year}{2010}{\natexlab{a}}) \bibinfo{pages}{306}.
\bibitem[{Bailey et~al.(2010{\natexlab{b}})}]{ref:fnal-2010-2}
\bibinfo{author}{J.~Bailey}, et~al.,
\newblock \bibinfo{title}{{$B \rightarrow D^* l \nu$ at zero recoil: an
  update}},
\newblock \bibinfo{journal}{PoS} \bibinfo{volume}{LATTICE 2010}
  (\bibinfo{year}{2010}{\natexlab{b}}) \bibinfo{pages}{311}.
\bibitem[{Bhattacharya and et~al.(1999)}]{ref:lanl-1999-1}
\bibinfo{author}{T.~Bhattacharya}, \bibinfo{author}{et~al.},
\newblock \bibinfo{title}{{Non-perturbative Renormalization Constants using
  Ward Identities}},
\newblock \bibinfo{journal}{Phys. Lett. B} \bibinfo{volume}{461}
  (\bibinfo{year}{1999}) \bibinfo{pages}{79}.
\bibitem[{{C. Bernard et al.}(2002)}]{milc-2002-1}
\bibinfo{author}{{C. Bernard et al.}},
\newblock \bibinfo{title}{{Lattice calculation of heavy light Decay constants
  with two flavors of dynamical quarks}},
\newblock \bibinfo{journal}{Phys. Rev. D} \bibinfo{volume}{66}
  (\bibinfo{year}{2002}) \bibinfo{pages}{094501}.
\bibitem[{Bernard et~al.(2011)Bernard, Detar, and
  Toussaint}]{milc-private-2011}
\bibinfo{author}{C.~Bernard}, \bibinfo{author}{C.~Detar},
  \bibinfo{author}{D.~Toussaint},
\newblock \bibinfo{journal}{private communication}  (\bibinfo{year}{2011}).
\bibitem[{Sivia and Skilling(2006)}]{sivia-2006}
\bibinfo{author}{D.~Sivia}, \bibinfo{author}{J.~Skilling},
  \bibinfo{title}{{Data Analaysis --- A Bayesian Tutorial}},
  \bibinfo{publisher}{Oxford University Press}, \bibinfo{edition}{second}
  edition, \bibinfo{year}{2006}.
\bibitem[{Lepage and et~al.(2002)}]{Lepage-2001}
\bibinfo{author}{G.~P. Lepage}, \bibinfo{author}{et~al.},
\newblock \bibinfo{title}{{Constrained curve fitting}},
\newblock \bibinfo{journal}{Nucl. Phys. Proc. Suppl.} \bibinfo{volume}{106}
  (\bibinfo{year}{2002}) \bibinfo{pages}{12--20}.
\bibitem[{Press et~al.(2007)Press, Teukosky, Vetterling, and Flannary}]{NR-7-4}
\bibinfo{author}{W.~Press}, \bibinfo{author}{S.~Teukosky},
  \bibinfo{author}{W.~Vetterling}, \bibinfo{author}{B.~Flannary},
  \bibinfo{title}{Numerical Recipes}, \bibinfo{publisher}{Cambridge University
  Press}, \bibinfo{address}{New York}, \bibinfo{edition}{third} edition,
  \bibinfo{year}{2007}. \bibinfo{note}{Chapter 7, section 4}.
\bibitem[{Hogg et~al.(2005)Hogg, McKean, and Craig}]{hogg-2005}
\bibinfo{author}{R.~V. Hogg}, \bibinfo{author}{J.~W. McKean},
  \bibinfo{author}{A.~T. Craig}, \bibinfo{title}{{Introduction to Mathematical
  Statistics}}, \bibinfo{publisher}{Pearson Prentice Hall},
  \bibinfo{edition}{sixth} edition, \bibinfo{year}{2005}.

\end{thebibliography}

\end{document}